\documentclass[fleqn,usenatbib]{mnras}

\usepackage{newtxtext,newtxmath}
\usepackage[T1]{fontenc}

\usepackage{graphicx}	
\usepackage{amsmath}	

\title[Super-Eddington growth of overmassive BHs]{Episodic super-Eddington accretion as a clue to Overmassive Black Holes in the early Universe}

\author[Trinca et al.]{
Alessandro Trinca$^{1,2,3,4}$\thanks{E-mail: atrinca@roe.ac.uk},
Rosa Valiante$^{3}$,
Raffaella Schneider$^{3,4,5,6}$,
Ignas Juodžbalis$^{7}$,
Roberto Maiolino$^{7,8}$,
\newauthor \:
Luca Graziani$^{3,4,5}$,
Alessandro Lupi$^{2,9}$,
Priyamvada Natarajan$^{10,11,12}$,
Marta Volonteri$^{13}$
and Tommaso Zana$^{3,4,5}$
\\\\
$^{1}$Institute for Astronomy, University of Edinburgh, Royal Observatory, Blackford Hill, Edinburgh EH9 3HJ, UK\\
$^{2}$Como Lake Center for Astrophysics, DiSAT, Università degli Studi dell'Insubria, via Valleggio 11, 22100 Como, Italy\\
$^{3}$INAF/Osservatorio Astronomico di Roma, Via Frascati 33, 00040 Monte Porzio Catone, Italy\\
$^{4}$INFN, Sezione Roma1, Dipartimento di Fisica, “Sapienza” Università di Roma, Piazzale Aldo Moro 2, 00185 Roma, Italy\\
$^{5}$Dipartimento di Fisica, “Sapienza” Università di Roma, Piazzale Aldo Moro 2, 00185 Roma, Italy\\
$^{6}$Sapienza School for Advanced Studies, Viale Regina Elena 291, 00161 Roma, Italy\\
$^{7}$Kavli Institute for Cosmology, University of Cambridge, Madingley Road, Cambridge CB3 0HA, UK\\
$^{8}$Cavendish Laboratory, University of Cambridge, 19 JJ Thomson Avenue, Cambridge CB3 0HE, UK\\
$^{9}$INFN, Sezione di Milano-Bicocca, Piazza della Scienza 3, I-20126 Milano, Italy\\
$^{10}$Department of Astronomy, Yale University, New Haven, CT 06511, USA\\
$^{11}$Department of Physics, Yale University, New Haven, CT 06511, USA\\
$^{12}$Yale Center for Astronomy \& Astrophysics, Yale University, New Haven, CT 06520, USA\\
$^{13}$Institut d’Astrophysique de Paris, Sorbonne Université, CNRS, UMR 7095, 98 bis bd Arago, 75014 Paris, France
}

\date{Accepted XXX. Received YYY; in original form ZZZ}

\pubyear{\the\year{}}

\begin{document}
\label{firstpage}
\pagerange{\pageref{firstpage}--\pageref{lastpage}}
\maketitle

\begin{abstract}
Early JWST observations are providing growing evidence for a ubiquitous population of accreting supermassive black holes (BHs) at high redshift, many of which appear overmassive compared to the empirically-derived local scaling relation between black hole mass and host galaxy stellar mass. In this study, we leverage predictions from the semi-analytical Cosmic Archaeology Tool (CAT) to reconstruct the evolutionary pathways for this overmassive BH population, investigating how they assemble over cosmic time and interact with their host galaxies. We find that the large $\rm M_{BH}-M_{star}$ ratios can be explained if light and heavy BH seeds grow by short, repeated episodes of super-Eddington accretion, triggered by major galaxy mergers. On average, we find that BH-galaxy co-evolution starts in earnest only at $z < 8$, when $\simeq 30\%$ of the final galaxy stellar mass has formed outside the massive black hole host. Our model suggests that super-Eddington bursts of accretion last between $0.5-3$ Myr, resulting in a duty cycle of $1-4\%$ for the target BH sample. The boost in luminosity of BHs undergoing super-Eddington accretion helps to explain the luminosity function of Active Galactic Nuclei observed by JWST. At the same time, a large population of these overmassive BHs is predicted to be inactive, with Eddington ratios $\lambda_{\rm Edd} < 0.05$, in agreement with recent observations.
\end{abstract}




\section{Introduction}
\label{sec:intro}
The James Webb Space Telescope (JWST) has opened a new observational window on the Active Galactic Nuclei (AGNs) population at high-redshift, allowing us to identify and study AGNs via optical and UV diagnostic diagrams even at low masses and low luminosities. Significant progress has been made through deep spectroscopic observations, which have revealed the presence of broad H$\alpha$ and H$\beta$ emission lines. Potential contamination due to the presence of outflows in these systems is disfavoured by the lack of detection of a broad counterpart in other bright emission lines, such as the forbidden transition of $\rm [OIII]5007$ \AA, clearly pointing towards the presence of a Broad Line Region \citep[BLR,][]{Maiolino2024z11} and hence an active AGN. These observations have revealed a previously unseen population of intermediate/low luminosity AGNs at $z > 4$, and out to $z \simeq 11$, with estimated black hole (BH) masses in the range $10^6$ to $10^8 \ \rm M_\odot$, and bolometric luminosities L$_{\rm bol} \sim (10^{43}-10^{46}) \rm \ erg/s$ \citep[e.g.][]{ubler2023, barro2024,  Maiolino2024bhs, harikane2023, larson2023, oesch2023, onoue2023, Ono2023, Yang2023, Leung2023}. These BH masses and luminosities are 2–3 orders of magnitude lower than those inferred for quasars at the same redshifts. In addition, the identification of narrow-line (i.e. type-2, as opposed to type-1 AGNs showing BLR emission) AGNs has been possible using UV and optical high-ionization emission lines \citep[e.g.][]{scholtz2024_AGNs,Mazzolari2025,Chisholm2024}. Of paramount importance is the discovery of three AGNs at $z \geq 10$ \citep{Maiolino2024z11, Bogdan2024, kovacs2024, napolitano2025} that are pushing our exploration closer to the seeding epoch, thereby enhancing our ability to constrain their origin  \citep[e.g.][]{natarajan2017unveiling, valiante2018observability, valiante2018statistics}. 

JWST has also revealed a previously unknown population of systems, dubbed “little red dots” \citep[LRDs; ][]{Matthee2024}, which consists of very compact sources with a steep, red, rest-frame optical continuum emission, presumably dominated by an obscured AGN component \citep[][and see also \citealt{fujimoto2022}]{Greene2024, labbe2023LRDs, furtak2023uncover}. This interpretation is supported by the detection of broad emission lines, although alternative explanations suggest that their properties could also be attributed entirely to stellar emission \citep{Baggen2024}. The fraction of LRDs hosting an AGN is still debated, with estimates ranging from $20 \%$ \citep{PerezGonzalez2024} to $80 \%$ \citep{Greene2024}, and it strongly depends on the detailed selection function. Conversely, the fraction of type-1 AGNs selected by JWST that exhibit LRD-like colors and slopes is reported to be between $10 \%$ and $30 \%$ \citep{Hainline2024,Maiolino2024bhs}.
While appearing fairly ubiquitous in the early Universe, the LRD population exhibits a strong evolution with redshift, reaching as high as $z \simeq 9$ before declining rapidly below $z \sim 3-4$ \citep{Kocevski2024}. While the apparent high-redshift limit is likely driven by current observational sensitivity, the sharp decrease toward lower redshifts is thought to reflect a physical evolution, suggesting that the extreme conditions required for the formation and/or appearance of LRDs become progressively rarer with cosmic time. This interpretation is further supported by the discovery of only a handful of potential LRD analogues in the local Universe \citep{Lin2026,Ji2026}, indicating that although similar systems can still exist at late times, they represent exceptionally uncommon environments. At the same time, growing observational evidence suggests that LRDs do not constitute a homogeneous class of objects, but instead span a broad range of spectral properties and continuum shapes, likely tracing different physical conditions and possibly distinct stages within a short-lived evolutionary sequence \citep{Perez-Gonzalez2026}. Nevertheless, a common characteristic emerging from recent spectroscopic observations is that LRDs are embedded within extremely dense, optically thick gaseous environments. This high-density gas is thought to be responsible for the prominent Balmer breaks observed in their spectra, while additional signatures of dense gas are revealed by complex Balmer line profiles, with deep absorption features and broad wings potentially shaped by electron-scattering contributions \citep{deGraaff2025,Ji2025,DEugenio2025,Rusakov2026,Torralba2026}.

These new JWST discoveries imply a very large number density of AGNs at $z \simeq 4  - 6$ \citep{Maiolino2024bhs, scholtz2024_AGNs, Greene2024,Kocevski2024, Akins2024}, which is 1 or 2 orders of magnitude larger than the extrapolation of the luminosity function probed by the quasar population \citep{niida2020}, and higher than the estimated density from deep X-ray observations \citep{giallongo2015, giallongo2019}.
Interestingly, a large fraction of these JWST-detected AGNs appears to be powered by BHs that are overmassive with respect to their host galaxy stellar component, with a BH-to-stellar mass ratio of even up to $\rm M_{\rm BH}/M_{\rm star}\sim 0.1-1$ \citep[][]{Maiolino2024bhs,Juodzbalis2024,Pacucci2024, harikane2023, ubler2023, furtak2024, kokorev2023, yue2024, Bogdan2024}. This is an offset that is at least two orders of magnitude higher than expected from the local $M_{\rm BH}-M_{\rm star}$ scaling relation \citep{KormendyHo2013,reines2015} and is challenging to reproduce in the context of cosmological simulations \citep{habouzit2022,bhowmick2024}.
This finding has been interpreted in some studies as a result of a large scatter in the BH-stellar mass relation, combined with expected selection effects \citep{Lauer2007}. Luminous and massive BHs, whose emission significantly dominates that of their host galaxy, are more likely to be detected in flux-limited surveys \citep[e.g.][]{volonteri2023, zhang2023, Li2024tipBHs,Sun2025,Silverman2025}. Systematic uncertainties in inferring BH masses from the observed broad emission-line properties may also lead to significant BH mass overestimation, and thus help mitigate the observed offset, particularly in the presence of super-Eddington accretion, anisotropic emission, or electron scattering-dominated line widths \citep{Lupi2024, king2024,Rusakov2026,Trinca2026}. Recent discoveries of overmassive BHs associated with low-luminosity AGNs, characterized by BH-to-stellar mass ratios closer to unity, provide tentative support for an intrinsic evolution of the $\rm M_{BH} - M_{star}$ relation with redshift \citep{Juodzbalis2024}. Furthermore, the finding that the $\rm M_{BH} - \sigma$ relation of these high-z AGNs aligns more closely with the local scaling suggests that the offset in the $\rm M_{BH} - M_{star}$ relation is not entirely driven by selection effects.
Even if selection biases play a significant role and we are detecting only a tail of the population, the observed overmassive black holes represent a collection of extraordinary objects that are challenging to explain within standard BH evolutionary scenarios.
Recent works have suggested that the observed population of massive BHs may be more easily explained if these systems descend from a population of \textit{heavy} seeds, formed already with masses of $10^4 - 10^6~ \rm M_\odot$ \citep{Pacucci2026, Begelman2026, Baggen2026}. However, for this scenario to account for the very high number density inferred for JWST broad-line emitters, the rare conditions required to form such massive seeds must be significantly relaxed relative to the predictions of detailed theoretical models and simulations to date \citep{habouzit2016}. Moreover, the recent detections of remarkably massive systems already in place at $z \gtrsim 8$ \citep{Taylor2025, Tripodi2025, Belladitta2026} further suggest that phases of rapid super-Eddington accretion might still be required in the early Universe, rather than massive seeding alone.

In this work, we aim to explore the implications of these findings for early BH formation and subsequent growth. In particular, we use a cosmological semi-analytical model, the Cosmic Archaeology Tool \citep[\texttt{CAT};][]{trinca2022,trinca2023bh}, to address the origin of overmassive BHs at $4 < z < 7$ and account for their large number densities.  

This model has already been extensively used to investigate how different black hole formation and accretion pathways shape the evolution of AGNs and galaxies across cosmic time, as well as to assess the detectability of faint MBHs at high-redshift with current and future facilities \citep{trinca2022,trinca2023bh}. 
Motivated by the unprecedented wealth of data recently provided by JWST surveys, we aim here to exploit the large statistical samples generated by the \texttt{CAT} semi-analytical model under different assumptions for early BH growth, in order to determine whether the peculiar properties of these newly observed MBHs can be more naturally reproduced within a specific evolutionary framework. In this way, we aim to place new constraints on the dominant growth mechanisms of the early MBH population.
In particular, we use the \texttt{CAT} model presented in \citet{trinca2022}, based on the same simulated galaxy and MBH samples, to distinguish between two evolutionary scenarios that were previously found to similarly reproduce the population of bright high-redshift AGNs observed prior to JWST, but which can now be tested in much greater detail thanks to the improved observational constraints and larger statistical samples extending toward the faint end of the AGN population. These scenarios correspond to: (i) a MBH population growing through relatively continuous, moderate accretion onto massive BH seeds, and (ii) MBHs experiencing episodic, highly efficient, and potentially super-Eddington, accretion onto both light and massive BH seeds.

We find that both the apparent “overmassiveness” and the high number density of the observed AGN population are reproduced significantly better in a scenario where BH seeds undergo short phases of super‑Eddington accretion.

The paper is organized as follows. The main features of our model are summarized in Section \ref{sec:model} while \texttt{CAT} predictions for the JWST overmassive BH population are detailed in Section~\ref{sec:overmassiveBHs}. We present the properties of the sample of synthetic overmassive BHs in Section~\ref{sec:sample}, showing the evolutionary connection to their host galaxies (Section \ref{sec:asynchGrowth}) and their predicted growth timescale (Section~\ref{sec:SEduration}). Specifically, we investigate the origin and fate of 
\texttt{CAT} BH candidates representative of the “dormant” BHs observed by \cite{Juodzbalis2024} at $z\sim 7$ (Section ~\ref{sec:DormantMBH}) and discuss the impact of our findings on the predicted AGN luminosity function (Section~\ref{sec:AGNlumFunc}). Finally, \texttt{CAT} results and their implications are critically discussed in Section~\ref{sec:discussion} and the main conclusions of this work are presented in Section~\ref{sec:conclusions}.

\section{Modeling the galaxy-black hole co-evolution}
\label{sec:model}

\texttt{CAT} is a semi-analytical model developed to characterize the formation of the first galaxies and their nuclear BHs, and to follow their subsequent co-evolution in a full cosmological context across cosmic time. 
The model adopts analytical parametrizations for a wide range of processes regulating galaxy evolution, based on well-established physical relations and/or results from detailed hydrodynamical studies. This framework allows the investigation of large statistical samples of the galaxy population and their evolution across cosmic time, resolving in detail a broad range of dynamical scales, from the first star-forming cosmic structures to massive and rare overdensities in the high-redshift Universe, while maintaining a contained computational cost. The free parameters adopted in the model have been calibrated in previous works \citep{trinca2022, trinca2023bh, trinca2024}, demonstrating that the model can reproduce simultaneously several key observables at high redshift, including the UV luminosity function, cosmic star formation rate density, reionization process, and BH mass functions.
In the following, we summarize the main features of the model, focusing on the modelling of early BH formation and growth, while referring the reader to the original papers for a more detailed description and for the calibration of the high-redshift galaxy population.

\subsection{Dark matter merger trees}

\texttt{CAT} relies on the \textsc{GALFORM} semi-analytic model \citep{cole2000, parkinson2008} to reconstruct the hierarchical merger history of a large sample of dark matter (DM) halos. This relies on a Monte Carlo algorithm based on a modified version of the Extended Press-Schechter (EPS) theory, explicitly tuned to match the conditional halo mass function extracted from the Millennium Simulation \citep{springel2005}. This calibration ensures that the statistical properties of the resulting merger trees are consistent with those of N-body simulations, both in their trends with halo mass and redshift, thereby mitigating the well-known limitations of EPS-based approaches in reproducing the abundance of most massive halos at high redshift \citep[see][for a detailed description of the model and validation tests]{parkinson2008}. We reconstruct the DM halo population between $z=4$ and $z=24$ through 110 DM merger trees, with a mass resolution corresponding to a minimum virial temperature of $T_{\rm vir} = 1200 \, \rm K$. This enables us to characterize the evolution of a large galaxy sample, representative of the global population, over the halo mass range $\rm M_{halo} \simeq 10^6 \, - \, 10^{14} \, \rm M_{\odot}$. \texttt{CAT} is therefore able to resolve the first episodes of star formation, taking place at $z>20$ inside molecular-cooling \textit{minihalos} ($T_{\rm vir} < 10^4 \, \rm K$) and to characterize in detail the formation of the first BH seeds that depend on the environmental properties of the host galaxies. 

\subsection{Baryonic evolution and star formation}
Following the virialization of dark matter halos, baryonic gas is accreted onto the interstellar medium (ISM) of newly formed systems through gradual infall, parametrized as $\dot{M}_{\rm inf} \propto (t/t_{\rm inf})^2 e^{(-t/t_{\rm inf})}$ \citep[see for details][and references therein]{valiante2011}. The infall timescale is assumed to scale with the halo free-fall time, $t_{\rm inf} = t_{\rm ff}/4$, while the normalization is chosen such that the total accreted gas mass asymptotically approaches the universal baryon fraction of the host halo. The fraction of gas mass within the ISM that is able to cool efficiently, and is therefore available for star formation ($M_{\rm gas}$), is determined by the interplay between gas cooling and dynamical timescales. In each galaxy, the star formation rate (SFR) is computed as:
\begin{equation}
     \textsc{SFR} = f_{\rm cool} M_{\rm gas} \epsilon_{\rm SF}/\tau_{\rm dyn},
    \label{eq:SFR}
\end{equation}
where $\rm \tau_{\rm dyn} \equiv [R_{\rm vir}^3 /(G M_{halo})]^{1/2}$ is the halo dynamical time, and $\epsilon_{\rm SF}$ is the star formation efficiency, which is a free parameter of the model that depends on the stellar population, and has been calibrated as detailed below. The factor $f_{\rm cool}$ quantifies the reduced cooling efficiency induced by Lyman–Werner (LW) radiation, in the $[11.2-13.6]$ eV energy band, which can photodissociate molecular hydrogen. It ranges between 0 and 1 and depends on the halo properties, being $=1$ only for atomic cooling halos ($T_{\rm vir} > 10^4 \, \rm K$). In minihalos, instead, $f_{\rm cool}$ is $<1$ and depends on the halo virial temperature, redshift, gas metallicity and intensity of the illuminating LW radiation \citep[][]{valiante2016, debennassuti2017, sassano2021}. The detailed evolution of $f_{\rm cool}$ as a function of these galaxy properties is determined by the comparison between the galaxy free-fall timescale and gas cooling timescale, where the latter is computed accounting for contributions from atomic and molecular hydrogen, metal fine-structure emission, and photoelectric emission from dust. A comprehensive treatment of the modelling can be found in Appendix A of \citet{valiante2016}.

The model also follows a self-consistent treatment of the chemical evolution of the ISM, including the coupled evolution of metals and dust in a two-phase medium, accounting for dust formation in supernova (SN) ejecta and in the atmospheres of asymptotic giant branch stars, dust destruction by interstellar shocks, and grain growth in molecular clouds. A detailed description of this chemical evolution framework is provided in \citet{valiante2011}.
We assume that Population~III (PopIII) stars form in pristine/metal poor halos, where the metallicity is $ \rm Z < Z_{\rm crit} = 10^{-3.8} \, Z_\odot$, according to a top-heavy distribution in the mass interval $10 \leq m_* \leq 300 \, \rm M_\odot$, parametrized by a Larson initial mass function \citep[IMF;][]{larson1998early}:
\begin{equation}
    \rm \Phi(m_*) = m_*^{\alpha-1} \, e^{-m_{*}/m_{ch}}
    \label{eq:larsonIMF}
\end{equation}
where $\alpha=-1.35$ and the characteristic mass is $\rm m_{ch} = 20 \, M_\odot$. As discussed in \citet{trinca2024}, the adopted Pop~III star formation efficiency is $\epsilon_{\rm SF, \, PopIII}=0.15$, to align with results of detailed hydrodynamical simulations of star formation in metal-poor clusters \citep{chon2022}.
This IMF is stochastically sampled as a function of the effective total stellar mass formed in each star formation episode \citep[][]{valiante2016, debennassuti2017}. 

In contrast, in halos enriched above the metallicity threshold, $ \rm Z > Z_{\rm crit}$, less massive, Population~II (PopII), stars are assumed to form according to Eq.~\ref{eq:larsonIMF} with masses in the range $0.1 \leq m_* \leq 100 \, \rm M_\odot$ and a characteristic mass $m_{ch}=0.35 \, \rm M_\odot$ \citep{debennassuti2014, debennassuti2017}. The efficiency of star formation for PopII host halos is assumed to be $\epsilon_{\rm SF, \, PopII}=0.05$. The adopted SF parameters have been shown to reproduce various properties of the high-redshift galaxy population, including the SFR density, stellar mass density, and UV luminosity function \citep{trinca2022,trinca2024}. 
While recent results suggest that a constant SF efficiency may be hard to reconcile with deep JWST observations at $z>10$ (\citealt{harikane2023}, but see also \citealt{Donnan2025}), \texttt{CAT} predictions for the evolution of the SFRD have been shown to closely match the values inferred from JWST observations up to $z \simeq  15$, when accounting for the observability limit of current surveys \citep{trinca2024}. This agreement is likely supported by the natural transition from PopIII to PopII dominated stellar populations that emerges from the model.

In addition to the star formation process modelled by Eq. \ref{eq:SFR}, one of the different evolutionary scenarios explored by our model also accounts for phases of enhanced star formation triggered by major galaxy mergers, as will be described in detail in Section \ref{sec:BHgrowth}.

\subsection{Black hole seed formation}

In this model, we follow the evolution of two seed BH populations: the \textit{light} seeds, remnants of the first generation of massive and metal-free stars (Population~III stars), and the \textit{heavy} seeds, originated through the so-called direct collapse scenario that requires specific environmental conditions \citep[see][for a recent review]{inayoshi2020}. 

In our model, PopIII stars forming in the mass interval $\rm 40 \, M_{\odot}< m_* < 140 \, M_\odot$ and $m_* > 260 \,\rm M_\odot$ are assumed to directly collapse into BHs of comparable mass \citep{heger2002}, after a time delay consistent with their stellar progenitor lifetime. The most massive among these Pop~III remnant BHs is then considered as a \textit{light} seed settling in the nucleus of its host galaxy thereafter \citep[see][for more details]{trinca2022, trinca2024}.

The population of \textit{heavy} seeds is instead characterized by a mass of $10^5 \, \rm M_\odot$ and is assumed to form in atomic-cooling halos (ACH, $T_{\rm vir}> 10^4 \, \rm K$) of pristine gas composition ($Z < Z_{\rm crit}$), to avoid fragmentation of the gas cloud, and illuminated by a LW flux, $\rm J_{LW}$, intense enough to dissociate H$_2$ molecules ($\rm J_{LW} > \, J_{crit}$), thus preventing star formation. Here we adopt a threshold value of $\rm J_{crit} = 300\times 10^{-21} \, \rm erg \, s^{-1} \, cm^{-2} \, Hz^{-1} \, sr^{-1}$ following \cite{trinca2022}. However, the value of this critical threshold, $\rm J_{crit}$, is highly uncertain \citep[see e.g.][for a discussion]{inayoshi2020}. We discussed the impact of different $\rm J_{crit}$ on heavy seed formation in previous studies, where we investigated the relative role of different seed populations in the assembly of $z\sim6$ SMBHs \citep{valiante2016, sassano2021}.

\subsection{Black hole mass growth}
\label{sec:BHgrowth}

After the formation of the initial BH seed, we follow the growth of these BHs through gas accretion and mergers across cosmic time. In \cite{trinca2022} we explored the imprints of different gas accretion regimes on the evolution of the global MBH population, as traced by their predicted 
mass and luminosity functions at different epochs. Here we consider two specific accretion scenarios, the \textit{Eddington-limited} (EL) and \textit{super-Eddington }(SE) models.

In the reference model, hereafter the Eddington-limited (EL) scenario, the gas accretion rate cannot exceed the Eddington rate and is calculated following the Bondi-Hoyle-Lyttleton (BHL) formula \citep{bondi1952,hoyle1941}:
\begin{equation}
    \dot{M}_{\rm BHL} = \alpha \, \frac{4 \pi G^2 M_{\rm BH}^2 \rho_{\rm gas}(r_A)}{c_{s}^3}
\label{eq:bondi}
\end{equation}
where $c_s$ is the sound speed, which is estimated
assuming a gas temperature $T_{\rm gas} = T_{\rm vir}$, and $\rho_{\rm gas}(r_A)$ the gas density evaluated at the radius of BH gravitational influence $r_A=2G M_{\rm BH} / c_s^{2}$. We approximate the gas density distribution as a singular isothermal sphere with a flat core:
\begin{equation}
\rho_{\rm gas}(r) =
\frac{\rho_{\rm norm}}{1 + \left(\frac{r}{r_{\rm core}}\right)^2}
\end{equation}
where $r_{\rm core} = 0.012 \, R_{\rm vir}$ and $\rho_{\rm norm}$ represents a normalization constant that ensures that, at each time step, the gas is distributed within the halo virial radius.
The factor $\alpha$ is a free parameter introduced in the original expression to account for the unresolved enhanced gas density in the inner regions around the BH. In fact, lack of resolution tends to strongly underestimate the actual BHL accretion rate in cosmological simulations \citep{dimatteo2012,schaye2015}. The same holds for semi-analytic models due to the inability to accurately estimate the gas density within the Bondi radius, which can be significantly enhanced compared to the average density inferred by assuming a smooth, cored gas profile at galactic scales. Following \citep{trinca2022}, in the EL scenario we assume $\alpha=90$, which enables us to reproduce the properties of the high redshift quasar population at $z \simeq 6-7$.

Although the BHL prescription is widely adopted in both semi-analytic models and hydrodynamical simulations, its applicability to the complex gas conditions expected in the first galaxies remains uncertain. These environments might be characterized by highly turbulent and inhomogeneous gas distributions, strong dynamical interactions, and rapidly varying inflow conditions, which challenge the assumptions of a smooth, stationary medium underlying the BHL formalism. We therefore also explored a scenario in which BH growth is temporarily enhanced by merger-driven gas inflows, supplementing the standard Bondi prescription.
In this alternative accretion paradigm, which we dub the super-Eddington (SE) scenario, gas accretion is enhanced during brief periods of time ($\sim 1-5$ \, \rm Myr) as a consequence of galaxy major mergers, defined as mergers amongst halos whose mass ratio is $\mu > 1/10$ \citep{tanaka2009}. 
The gravitational torques generated during these catastrophic events are expected to induce an efficient loss of angular momentum in the gas, triggering strong inflows toward the nuclear regions \citep{Hernquist89,Barnes91}. This is expected to drive an enhancement both in AGN activity and star formation within the galaxy, as supported by detailed numerical simulations \citep{dimatteo2005,Li2007,capelo2015}. As a result, the rate at which the gas is accreted onto a BH can be higher than the Eddington limit. Since the model does not directly follow the evolution of the gas angular momentum, we model the BH accretion rate in this phase as: 
\begin{equation}
    \dot{M}_{\rm BH} = \frac{\epsilon_{\rm BH} \, M_{\rm gas}}{\tau_{\rm accr}} 
    \label{eq:SErate}
\end{equation}
where $M_{\rm gas}$ is the gas mass inside the newly formed galaxy, $\epsilon_{\rm BH}=0.017$ and $\tau_{\rm accr}=10 \, \rm Myr$ are the assumed BH accretion efficiency and timescale, respectively. The chosen parameter values represent a conservative estimate, which is comparable to the maximum dynamical timescale of merging galaxy bulges found in the simulations by \citet{pezzulli2016}.
At the same time, an increased gas content in the central galactic regions is also expected to boost the galaxy star formation. For this reason, we also induce an enhancement of the star formation rate (SFR) of the newly formed galaxy, with respect to the prescription presented in Eq.~\ref{eq:SFR}, assuming a star formation timescale consistent with the BH accretion timescale $\tau_{\rm dyn,SF}=\tau_{\rm accr}$, so that
\begin{equation}
\textsc{SFR}_{\rm burst}=\frac{\epsilon_{\rm SF} M_{\rm gas}}{\rm \tau_{dyn,SF}}
\end{equation}
with $\epsilon_{SF}=0.05$ being the same star formation efficiency adopted in the EL model. This should be regarded as the simplifying assumption that the merger-driven gas inflow drives both the enhanced BH accretion and a compact nuclear starburst, so that the two processes evolve on a comparable dynamical timescale. We refer the reader to \citet[][]{pezzulli2016} and \citet{trinca2022} for a more in-depth description of the assumed set of free parameters and the model calibration.
In the SE scenario, the enhanced accretion is sustained as long as one of the two following conditions is met: the elapsed time interval reaches a value $\Delta t = \tau_{\rm dyn}/100$, 
or the gas-to-BH mass ratio drops below $M_{\rm gas}/M_{\rm BH}<10$.
Thereafter, gas accretion and star formation return to their quiescent modes, with star formation described by Eq. \ref{eq:SFR} and BH accretion proceeding at the BHL rate (Eq.~\ref{eq:bondi}). It is important to note, however, that in the super-Eddington model we assume a conservative boosting factor of $\alpha = 1$.
A detailed discussion of the motivation and caveats of the adopted model for merger-driven inflows and related phases of enhanced BH accretion will be provided in Section \ref{sec:discussion}.

Finally, nuclear BHs can also grow through mergers following the coalescences of their host galaxies.
In major mergers, we assume that the two nuclear BHs efficiently sink into the centre of the newly formed galaxy instantaneously merging, within the typical time step of the simulation $\Delta t \sim 0.5-4$\,Myr, depending on redshift (cosmic time). Conversely, in minor galaxy mergers ($\mu < 1/10$), only the more massive of the two BHs is considered as the nuclear engine of the resulting galaxy while the subsequent evolution of the less massive one is no longer followed. The timescales for binary BH mergers are uncertain and we note here that our assumption of instantaneous merging is optimistic. 

\subsection{Stellar and AGN feedback}

In each galaxy, the time evolution of the available gas mass is affected by radiative and mechanical feedback. 

In addition to the photodissociating LW radiation regulating the efficiency of star formation, we account for the effect of photo-heating feedback inducing an increase in the gas temperature that may suppress star formation. This occurs in halos with virial temperatures below the temperature of the intergalactic medium, at a given redshift \citep{valiante2016}.

In \texttt{CAT} we assume that the coupling of the gas with the energy released during SN explosions and the BH accretion process may deplete the reservoir available to star formation and BH accretion within a galaxy, launching powerful outflows that can extend to galaxy-scales.  
We describe the gas outflow rate due to the combined effect of SN and AGN mechanical feedback as $\dot{M}_{\rm ej}=\dot{M}_{\rm ej,SN}+\dot{M}_{\rm ej,AGN}$, where:
\begin{equation}
    \dot{M}_{\rm ej,SN} = 2 E_{\rm SN} \epsilon_{\rm w, SN} R_{\rm SN}(t)/v_{\rm e}^2
    \label{eq:SNfeedback}
\end{equation}
and
\begin{equation}
    \dot{M}_{\rm ej,AGN} = 2 \epsilon_{\rm w, AGN} \epsilon_{\rm r}(t) \dot{M}_{\rm accr} c^2/v_{\rm e}^2.
    \label{eq:AGNfeedback}
\end{equation}
In these equations, $v_{\rm e}=\sqrt{2 G M/R_{\rm vir}}$ is the escape velocity of the halo, $E_{\rm SN}$ is the explosion energy per SN, $R_{\rm SN} (t)$ is the total SN explosion rate and $\dot{M}_{accr}$ is the gas accretion rate, computed as in Eq.~\ref{eq:bondi} (\ref{eq:SErate}) in the EL (SE) model.
The AGN radiative efficiency, $\epsilon_{\rm r}$, is fixed to the typical value of 0.1 \citep{shakura1973} in the EL model, while in the SE scenario it can vary according to the \citet{madau2014a} fitting formula for a BH spin of a=0.572 \citep[see][for further details]{trinca2022}.
Finally, the SN- and AGN-driven wind efficiencies, $\epsilon_{\rm w, SN}$ and $\epsilon_{\rm w, AGN}$, are free parameters of the model and are assumed to be $1.6\times 10^{-3}$ and $2.5\times 10^{-3}$, respectively for both the EL and SE models \citep{trinca2022}.
We do not include specific AGN feedback during super-Eddington phases, which are expected to be associated with the emission of powerful bipolar jets from the accreting BH. Although the impact of jets remains debated, they might play a significant role in limiting BH growth, especially in smaller galaxies \citep{regan2019}. However, given the relatively large value assumed in \texttt{CAT} for the coupling parameter $\epsilon_{w, AGN}$, episodes of SE growth are very efficient in depleting the gas reservoir in less massive systems.

\section{A population of Overmassive Black Holes}\label{sec:overmassiveBHs}

\begin{figure*}
\centering
\includegraphics[width=0.7\hsize]{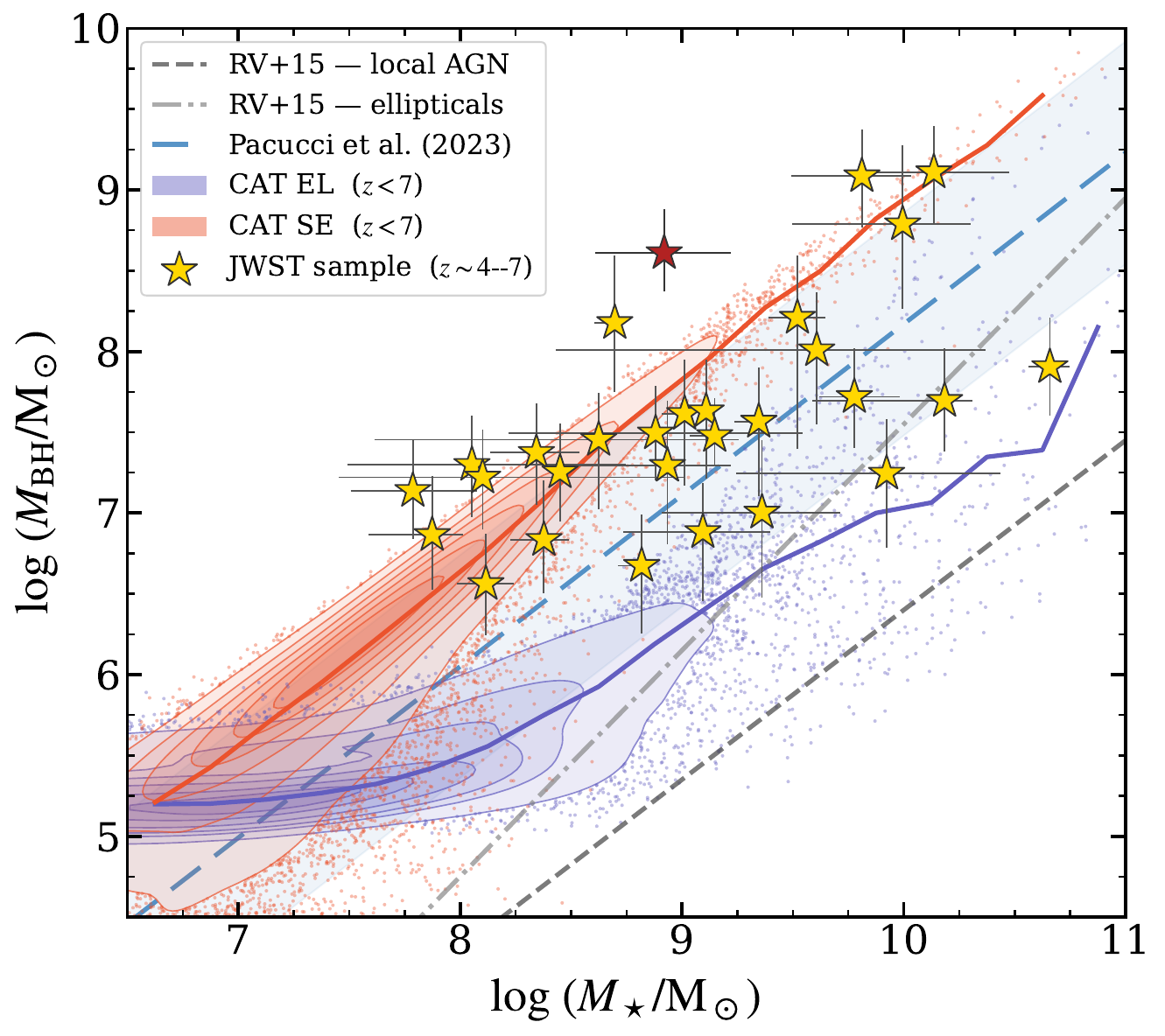}
\caption{Scaling relation between black hole mass and host-galaxy stellar mass. Purple and red data points (with the corresponding density contours and solid median lines) represent the BH population predicted at $4<z<7$ by the \texttt{CAT} Eddington-limited (EL) and super-Eddington (SE) accretion models, respectively. Model predictions are compared with a sample of overmassive BHs detected in early JWST observations at high redshift ($4<z<7$; yellow stars). The dormant black hole at $z=6.67$ identified in the JADES system GN-1001830, analysed in Section~\ref{sec:DormantMBH}, is highlighted as a red star. Grey dashed and dash-dotted lines show the local scaling relations of \citet{reines2015} for local AGN (active galaxies) and ellipticals (passive galaxies), respectively. The blue dashed line and shaded region show the high-redshift scaling relation proposed by \citet{Pacucci2023}, together with its inferred scatter.}
\label{fig:MbhMstarObs}
\end{figure*}

In Figure~\ref{fig:MbhMstarObs} we compare the BH vs stellar mass relation predicted by the EL and SE models with the sample of AGNs detected by JWST in low-mass galaxies at redshift $4 < z < 7$ (sampled from \citealt{harikane2023, ubler2023, Maiolino2024bhs, stone2023, furtak2024, kokorev2023, yue2024}, see further details in \citealt{Mezcua2024}), aiming to test which evolutionary scenario describes better their estimated properties.
We find that the observed massive BH sample is better reproduced by the SE model, where the simulated systems closely match the region of the $\rm M_{BH}-M_{star}$ plane occupied by observational data, although the treatment of BH accretion and feedback in the super-Eddington regime in \texttt{CAT} is relatively simplified (see Section \ref{sec:model}).
In the EL model, the predicted $M_{\rm BH}/M_{\star}$ ratio is instead significantly lower than observed, more consistent with the local scaling relations for active and passive galaxies (dashed and dot-dashed lines, respectively), and the distribution of systems shows a larger scatter, especially at $M_{\rm star}>10^{8.5}$ M$_\odot$.

Further, we note that the flattening of the BH mass distribution in the $M_{\rm BH}\sim 10^5-10^6 \, \rm M_\odot$ range is a consequence of the BH accretion prescription adopted in the EL model. In this regime, the growth of light seeds is inefficient and, therefore, the MBH population originates preferentially from heavy seeds \citep{trinca2022}. Conversely, in the SE scenario a fraction of the light seed population can grow efficiently already at early times driven by repeated bursts of super-Eddington accretion, resulting in a wider distribution at the low-mass end.
 
\texttt{CAT} predictions suggest therefore that the overmassive systems detected by JWST 
are more consistent with a population of objects where the central BHs experienced accelerated growth through repeated episodes of super-Eddington accretion driven by their host galaxy dynamical interactions. The larger $\rm M_{\rm BH}/M_{*}$ ratios predicted in the SE model are a consequence of the response of BH growth and stellar mass assembly to merger-driven gas inflows. During major mergers, both BH accretion and star formation are enhanced in the remnant (post-merger) galaxy, and assumed to proceed on a comparable timescale. This drives the progressive increase in $\rm M_{\rm BH}/M_{*}$ relative to what is expected during more quiescent phases, in which stellar assembly dominates. The associated AGN feedback compounds this effect by depleting the remnant's gas reservoir, thereby suppressing subsequent star formation over an extended period of time. This effect is particularly pronounced at high redshift, where major mergers are more frequent and the shallower potential wells of low-mass halos make feedback-driven gas removal more effective.

As shown in Figure~\ref{fig:MbhMstarObs}, the SE model predicts a maximum BH-to-stellar mass ratio of $\sim 0.1$ for the galaxy population at $4 \lesssim z \lesssim 7$, reflecting the simultaneous enhancement of star formation and BH accretion rates during major galaxy mergers (see Section~\ref{sec:BHgrowth}), combined with the assumption that enhanced BH accretion can not be sustained in gas-poor environments, where $M_{\rm gas}/M_{\rm BH} < 10$ \citep[see][for details]{trinca2022}.
Therefore, within our SE scenario, the existence of AGNs with extreme $M_{\rm BH}/M_{\rm star}$ ratios suggests that, while merger-driven gas inflows enhance both BH accretion and star formation, their cumulative impact leads to a rise in the BH-to-stellar mass ratio compared to standard quiescent evolution.
To further characterize the SE scenario, we present in Appendix \ref{sec:appendix} additional diagnostics of the CAT massive BH population. In particular, we show the predicted black hole mass function at $ z \sim 5$, and test its consistency against the recent JWST constraints probing the distribution in the low-mass regime, down to $M_{\rm BH} \gtrsim  10^6 ~\rm M_\odot$  \citep{Taylor2025b, Geris2026}.

\begin{figure}
\centering
\includegraphics[width=\hsize]{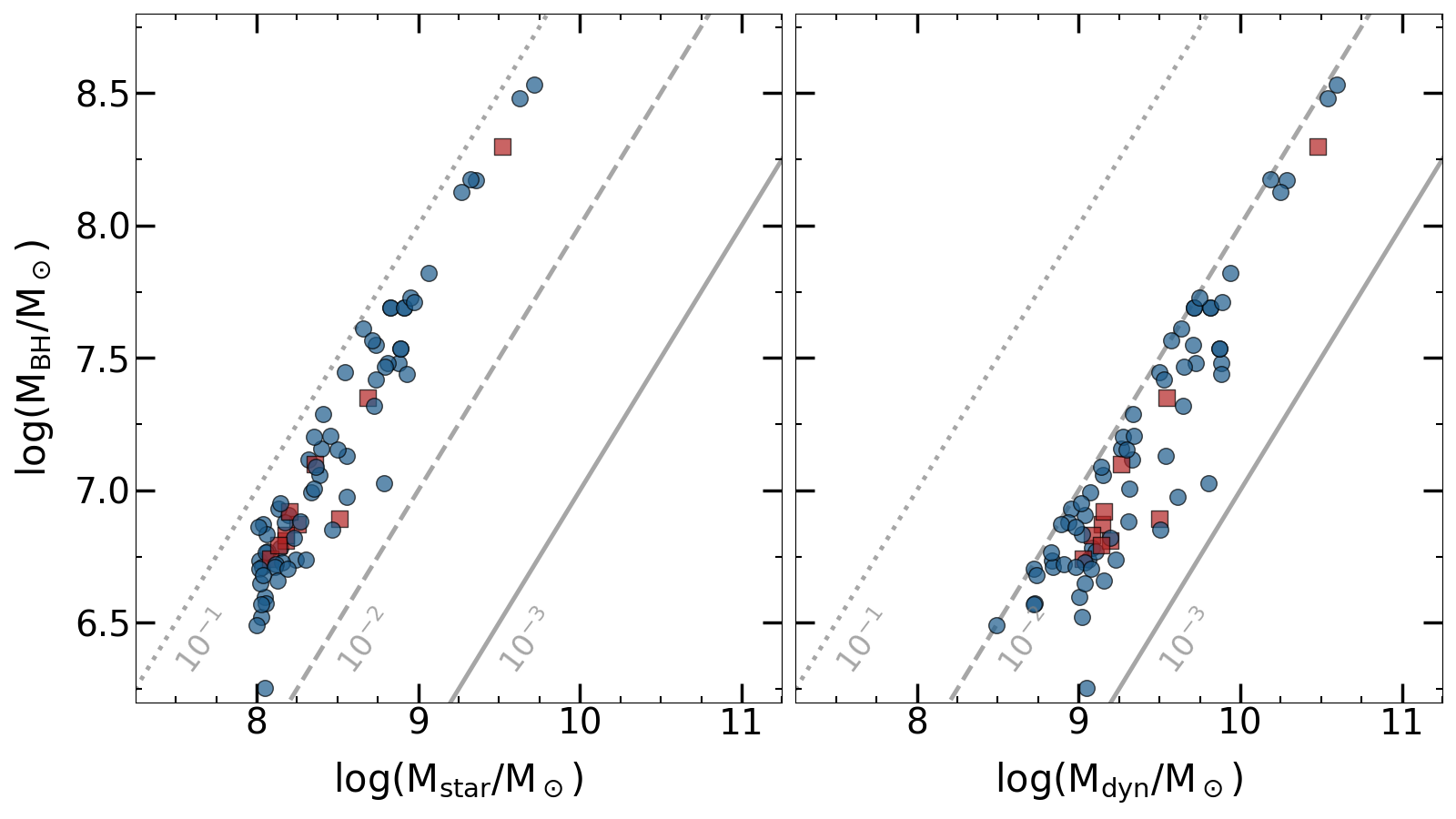}
\includegraphics[width=\hsize]{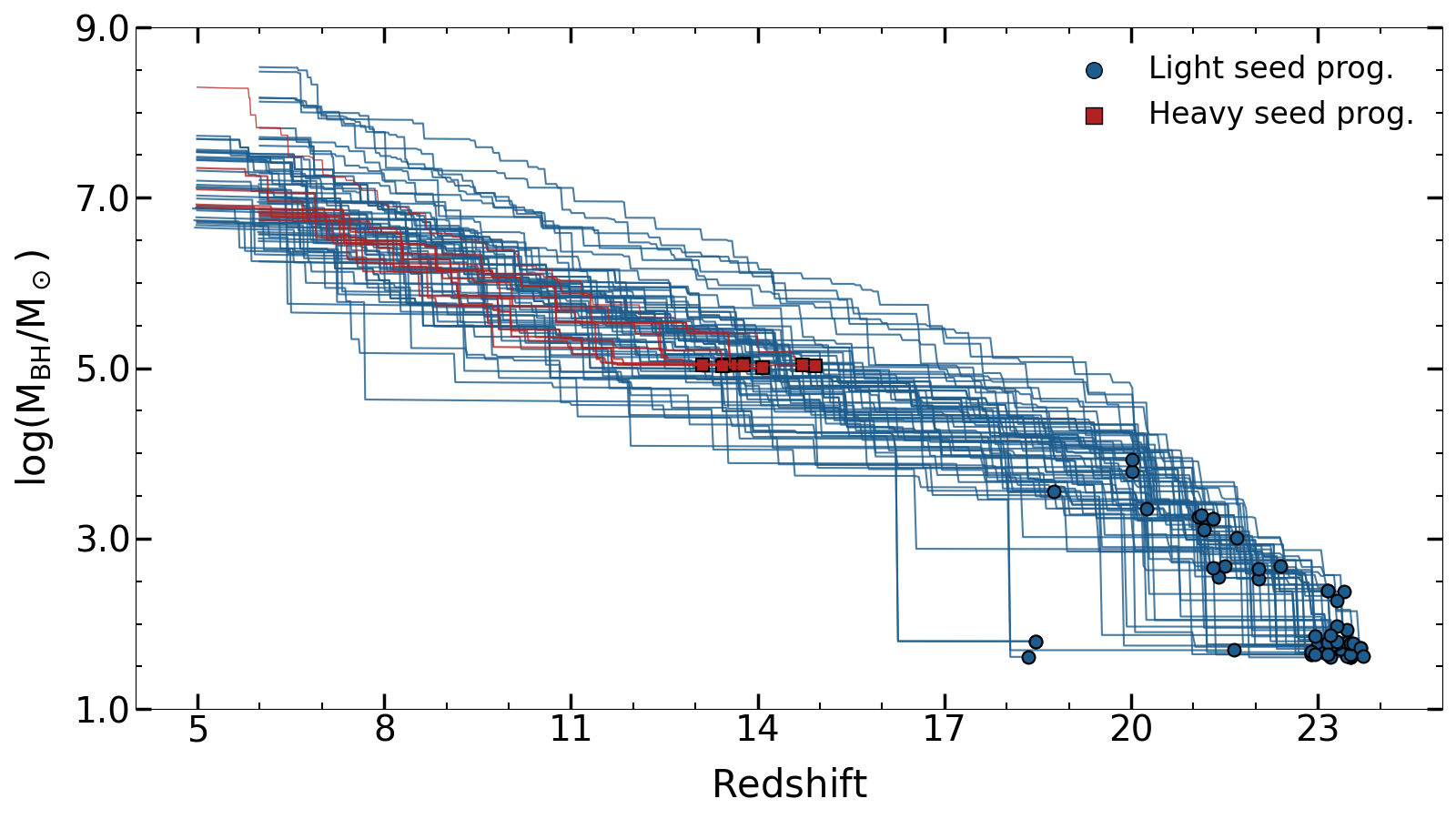}
\caption{Top panels: the \texttt{CAT} synthetic candidates at $z \sim 5$ and $z \sim 6$ in the $\rm M_{BH}-M_{star}$ and $\rm M_{BH}-M_{dyn}$ planes. Diagonal lines show constant ratio values of $10^{-1}, \, 10^{-2}$ and $10^{-3}$. Bottom panel: the initial seed masses and the subsequent BH evolutionary tracks for the selected systems. BHs originating from light and heavy seeds are represented, respectively, as blue dots and red squares.}
\label{fig:selectedPop}
\end{figure}

\subsection{Properties of overmassive BHs and their hosts}
\label{sec:sample}

In what follows, we analyse a sample of simulated systems in the SE model that best reproduce the observed AGN population at $z>4$. To do so, we select systems from the SE simulated catalogue at $z=5$ and $z=6$ with an expected number density large enough to result in at least 1 object in the area covered by the JADES-Medium survey, equal to $190 \rm \,\, arcmin^2$. Additionally, we require the systems to host a nuclear BH significantly more massive than expected from the local scaling with the host galaxy's stellar mass ($\rm log( M_{BH}/M_{star}) > -2$), but closer to the expected relation with the estimated galaxy dynamical mass ($\rm log( M_{BH}/M_{dyn}) < -2$). With this choice, we ensure to focus on a population that resembles more closely the BHs observed by \citet{Maiolino2024bhs} and \citet{Juodzbalis2024}.
We do not apply here any selection based on the luminosity of the investigated systems. This would require accounting for several critical factors affecting their observability, including potential obscuration, the impact of highly episodic accretion (see Sec. \ref{sec:SEduration}), and the detectability of AGN signatures such as broad emission lines. Our primary goal here is instead to investigate the nature and potential evolutionary pathway of the observed population.

The selected sample is shown in the upper panels of Figure~\ref{fig:selectedPop}, where we also indicate if the MBH originates from a light (blue) or a heavy (red) initial seed. Among the total population of $78$ systems, only $10$ arise from heavy seeds ($\simeq 13\%$). This we believe reflects the lower efficiency of the formation of heavy seeds, which requires more restrictive environmental conditions, as described in Section \ref{sec:model} \citep[see also][]{lodato2007,valiante2016,sassano2021}; this relative proportion also depends on the adopted prescription for SE growth, which allows light seeds to rapidly increase their mass, making them equally competitive in driving the growth of massive BHs, and their significantly larger number density is reflected in the selected MBH sample shown in Figure~\ref{fig:selectedPop}. The figure shows that the properties of the host galaxies and their nuclear BHs do not depend on the initial seed mass by this epoch in cosmic time. In fact, we find no significant differences between systems descending from light and heavy seeds in the BH-stellar mass and BH-dynamical mass planes. This is also confirmed by the BH evolutionary tracks shown in the bottom panel of the figure. The tracks show that, by the time the first heavy seeds form, in the redshift interval $z \sim 13- 15$, several light seed descendants have been able to grow to masses comparable to, or even higher than, those of the heavy seeds \citep[see also][]{pezzulli2016}, and no peculiar features in the BH evolutionary tracks allow one to discern the evolution of the two populations thereafter.

It is worth noting that, while in \texttt{CAT} we find direct collapse BHs formed in pristine atomic cooling halos even at earlier times in a cosmological environment \citep[see for instance][]{trinca2022}, the high photodissociating background flux required to prevent fragmentation could be provided only in very large overdensities, which are extremely rare at $z>15$. Consequently, these systems would not match the estimated number density of JWST-detected MBHs and are excluded from our sample selection.
Assuming a more widespread mechanism of formation for heavy seeds at early times, or a less efficient growth through accretion of lighter BH seeds \citep{smith2018} would naturally shift the obtained proportion in favour of the former population.

\subsection{Decoupled galaxy - BH growth}
\label{sec:asynchGrowth}
In this Section, we explore the evolution in time of the stellar component for the selected sample of galaxies hosting overmassive BHs. To assess how the stellar mass and BH mass co-evolve along the cosmological assembly of each system, and therefore how they may influence each other, we distinguish between star formation occurring \textit{in situ}, i.e. within the same halo as the most massive BH progenitor, and \textit{ex situ}, i.e. in all other progenitor galaxies of the final system.

Note that, since the most massive BH progenitor seed can form in a satellite galaxy, as is typically the case for direct-collapse black holes, our definition of \textit{in situ} SF does not necessarily correspond to the star formation occurring in the main halo progenitor of the final galaxy, especially at early times. Instead, it refers to the stellar mass formed within the host halo of the most massive BH progenitor, i.e. the component that directly co-evolves with it.
 
\begin{figure}
\centering
\includegraphics[width=\hsize]{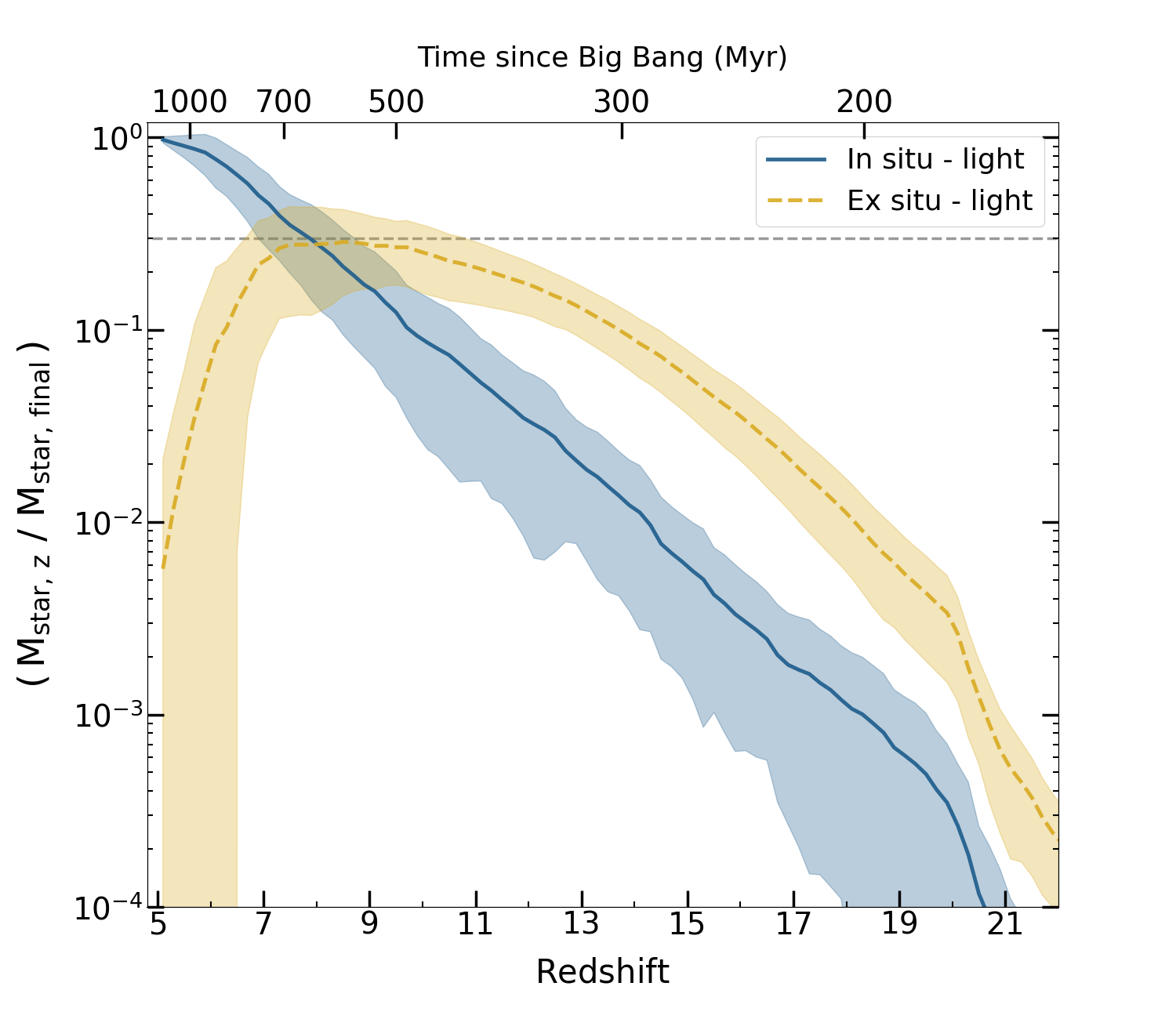}
\includegraphics[width=\hsize]{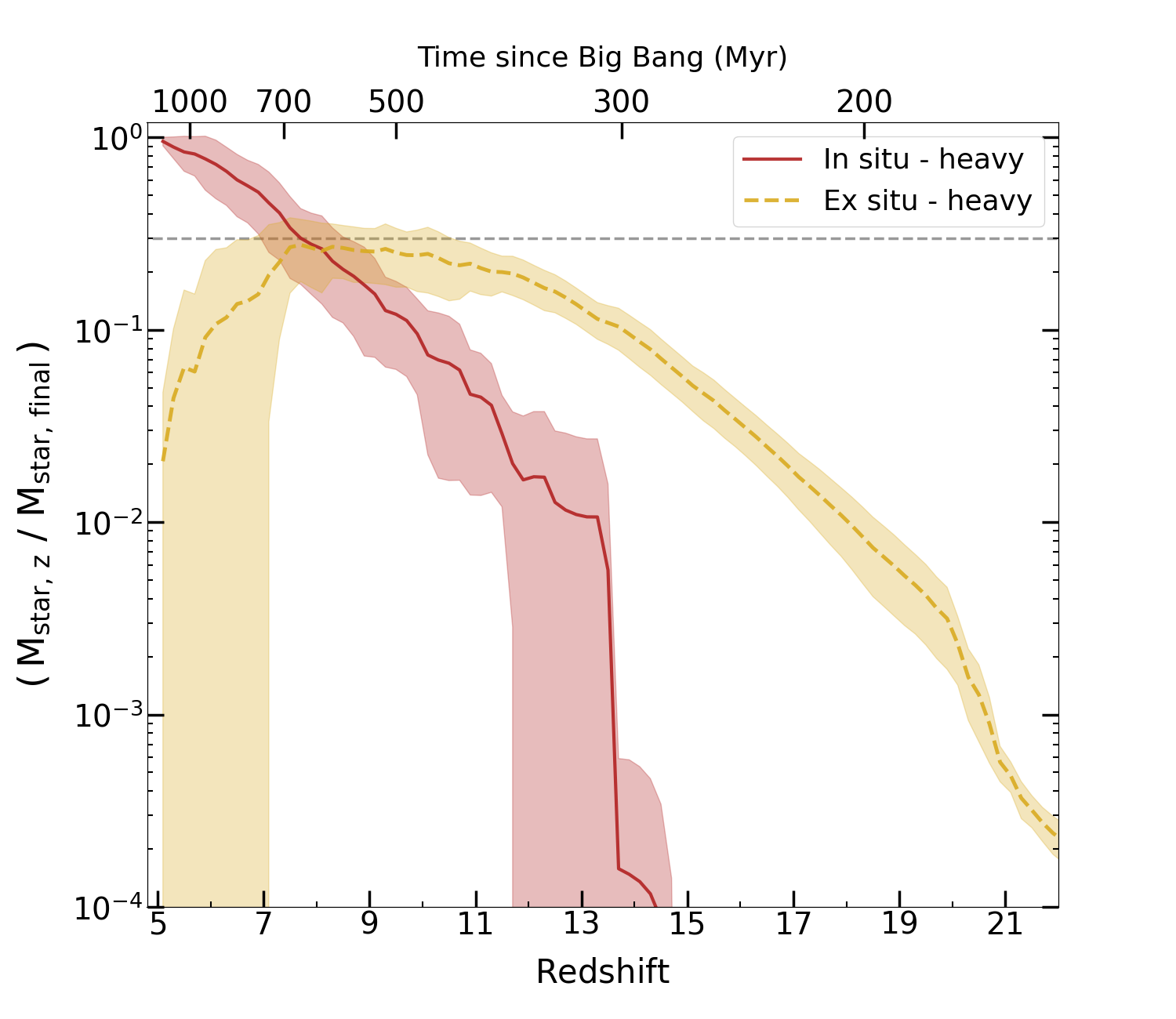}
\caption{Time evolution of the fraction of the final stellar mass assembled in galaxies hosting MBHs descended from light (top panel) and heavy (bottom panel) BH seeds. The solid and dashed lines represent the average stellar mass fractions formed \textit{in situ} (i.e. within the same galaxy hosting the final BH) and \textit{ex situ} (i.e. in other galactic progenitors), respectively. Shaded regions show the 1 $\sigma$ standard deviation around the average evolutionary history. The corresponding cosmological time is reported on the upper axis. The horizontal dashed gray line indicates $30 \%$ of the final galaxy stellar mass.}
\label{fig:AsGrowth}
\end{figure}

Figure~\ref{fig:AsGrowth} shows the average fraction of the galaxy's final stellar mass that has formed by a given redshift, distinguishing between \textit{in situ} and \textit{ex situ} star formation. The results are presented separately for MBH populations originating from \textit{light} seeds (upper panel) and \textit{heavy} seeds (lower panel).
In both panels, the shaded regions represent the $\rm 1 \, \sigma$ dispersion in the star formation histories of galaxies in the selected sample. As shown, the deviation from the average \textit{in situ} evolution is larger at early cosmic epochs, as a consequence of the different seed formation times, and becomes smaller at lower redshifts ($z\lesssim 8$), when most of the massive BH progenitors converge inside the main galaxy branch. This suggests that the selected overmassive systems share a similar assembly history in the last few hundred Myr of their evolution, while their early histories were more diverse in terms of both seed mass and growth mode (EL or SE).

We find that at $z \gtrsim 8$ a significant fraction of the stellar mass, $\sim 30\%$ (on average), grows in progenitor galaxies which do not co-evolve with the MBH, independently of whether the MBH descends from a light or a heavy seed. 
This behaviour reflects a temporary decoupling between BH growth and galaxy assembly. While the progenitor hosting the most massive BH experiences repeated merger-triggered super-Eddington episodes and strong AGN feedback, other branches of the merger tree continue forming stars more efficiently. As a result, a non-negligible fraction of the stellar mass formed in this early phase is assembled ex-situ (with respect to the MBH progenitor) and only later incorporated into the MBH host galaxy through hierarchical merging.
At $z \sim 8$, the systems enter a distinct evolutionary phase, during which the actual BH-galaxy co-evolution appears to begin. In this later stage, the galaxy assembles the majority of its stellar mass, with most systems building up over $70 \%$ of their final $M_{\rm star}$. 

These results reflect the efficiency of BH accretion and feedback in the early stages of BH seed growth.
At $z \gtrsim 8$, the primary BH progenitor evolves rapidly within halos that are undergoing frequent dynamical interactions due to their characteristic hierarchical assembly history, in which repeated episodes of SE accretion are triggered, effectively suppressing star formation in their host galaxies.

At $z < 8$, the central BH continues to grow efficiently, but its feedback effect on the host is weaker. This is due to two factors. On the one hand, the occurrence of major merger events - and thus of strong BH-driven feedback phases - decreases at lower redshift \citep{Ferreira2020, Puskas2025}, so that the host galaxy star formation becomes more efficient and the stellar mass grows faster. On the other hand, as the host DM halos grow in mass, deepening their gravitational potential well, gas outflows become less effective at depleting the galaxy's gas reservoir.

\subsection{Duration of Super-Eddington bursts}
\label{sec:SEduration} 
\begin{figure}
\centering
\includegraphics[width=\hsize]{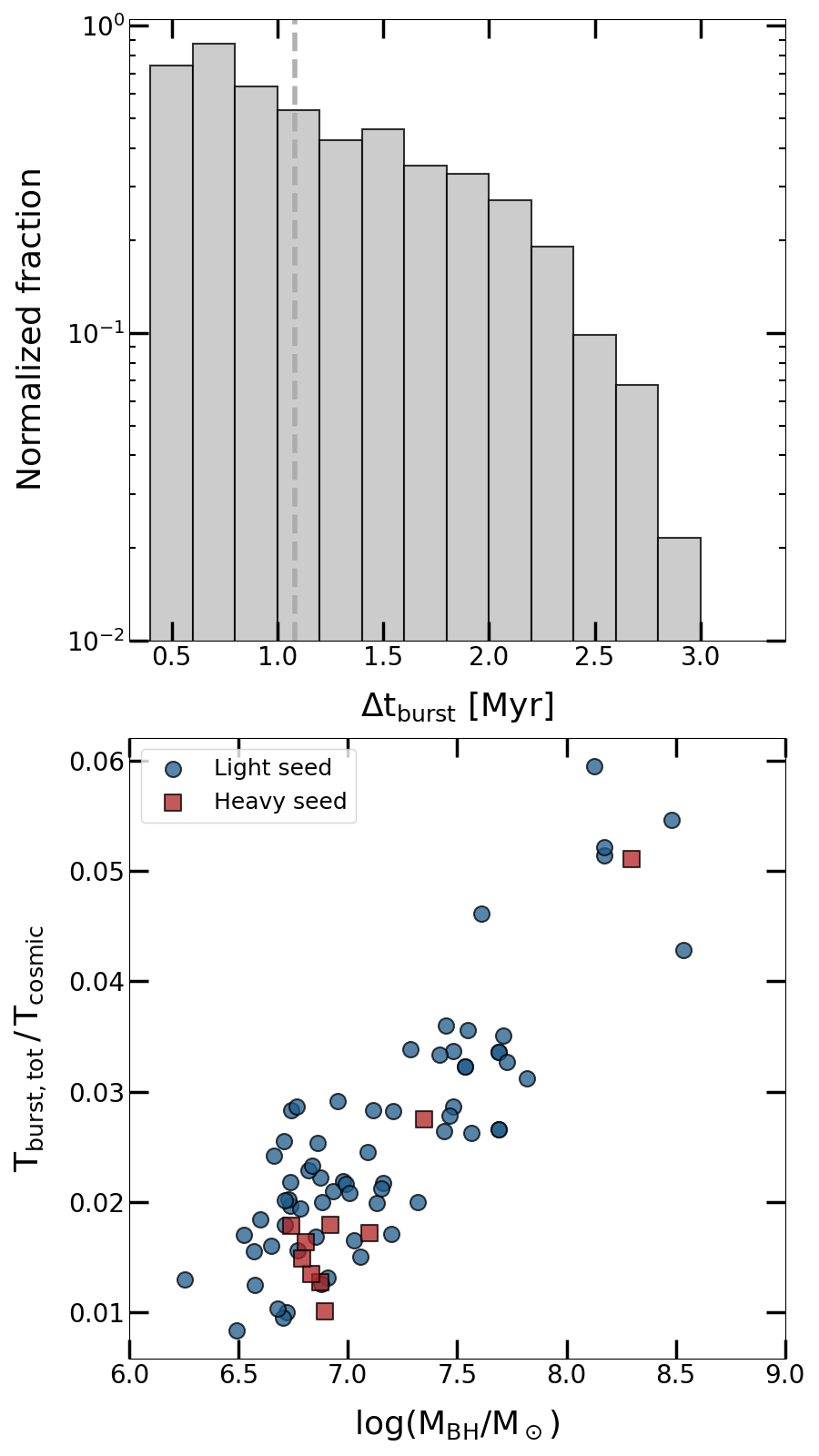}
\caption{\textit{Top panel}: distribution of the time duration of super-Eddington accretion bursts for the selected SMBH population shown in Figure \ref{fig:selectedPop}. The grey vertical line shows the median of the distribution. \textit{Bottom panel}: total fraction of the cosmological time spent in a burst phase as a function of the final BH mass. The points are color-coded according to the seeding scenario as in Figure \ref{fig:selectedPop}.}
\label{fig:DTburst}
\end{figure}

It is interesting to explore the typical duration and frequency of super-Eddington accretion episodes that the selected overmassive BHs experience during their evolution. This is shown in Figure \ref{fig:DTburst}. The distribution of the duration of single SE accretion bursts is presented in the upper panel, which shows that the typical values are very short, and range between $0.5-3$ Myrs, with a median value of $\Delta t_{\rm burst} \simeq 1$ Myr. This is consistent with the typical duration of SE phases suggested by high-resolution hydrodynamical simulations (e.g. \citealt[][]{sassano2023, massonneau2023, Lupi2024SE, Gordon2024}, see also the discussion in Section \ref{sec:discussion}).

The bottom panel of Figure \ref{fig:DTburst} instead shows the total AGN duty cycle of the selected sample, defined as the fraction of cosmological time spent in a phase of SE accretion during the repeated episodes of growth. We find SE duty cycles of a few percent, which increase with BH mass ranging from $0.5 \%$ for $ \rm M_{BH} \sim 10^{6.5} \, M_\odot$ to $6 \%$ for $ \rm M_{BH} \sim 10^{8.5} \, M_\odot$. This relatively small duty cycle implies that in the SE model most of the BHs are largely {\it dormant} and only a small fraction can be observed in their active phase \citep[see ][]{Juodzbalis2024}. This issue will be discussed further in Section \ref{sec:DormantMBH}.

\section{Massive dormant Black Holes}
\label{sec:DormantMBH}

\begin{figure*}
 \centering
    \hspace{-2.0cm}
    \includegraphics[width=0.7\textwidth]{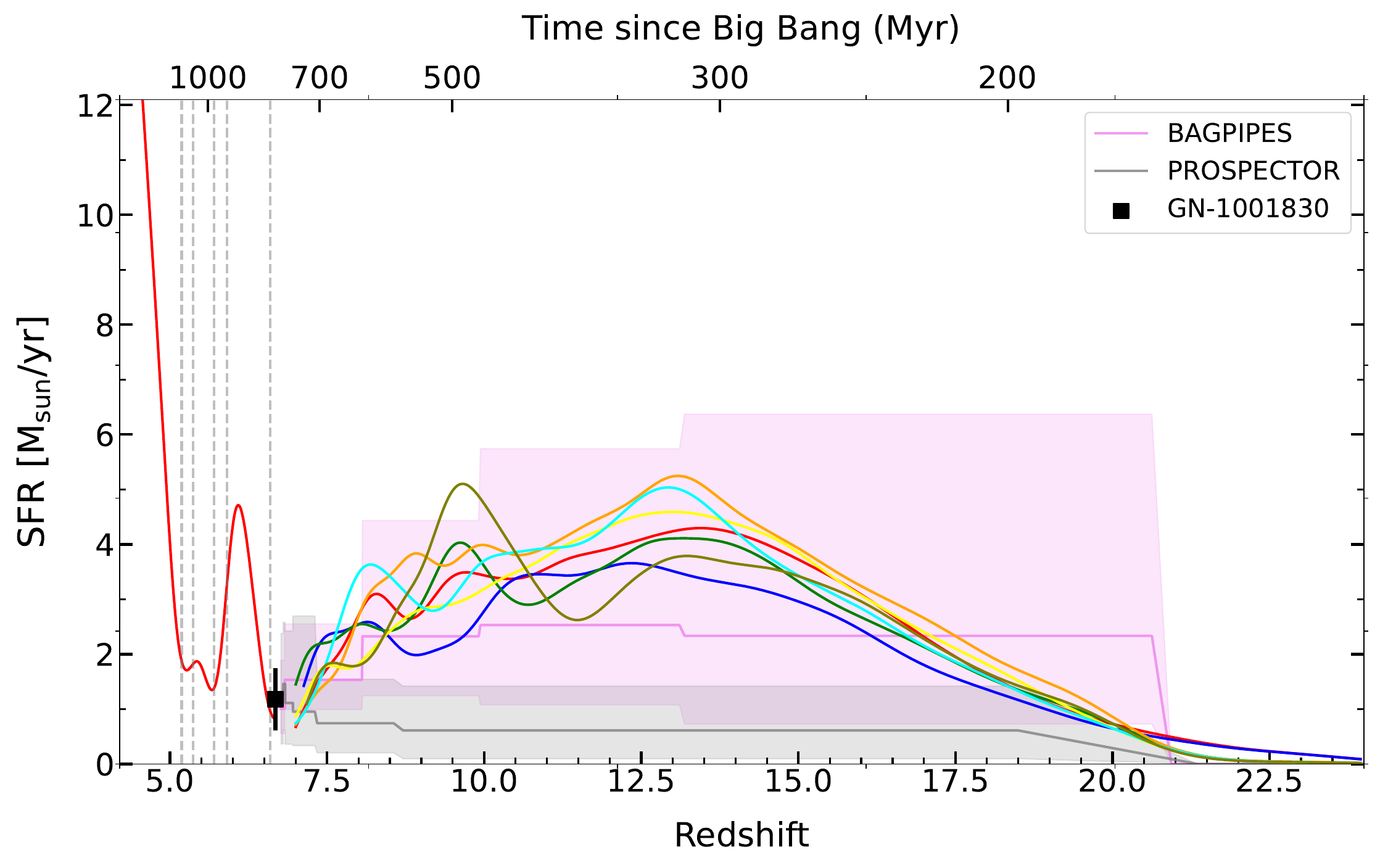}
    \includegraphics[width=0.7\textwidth]{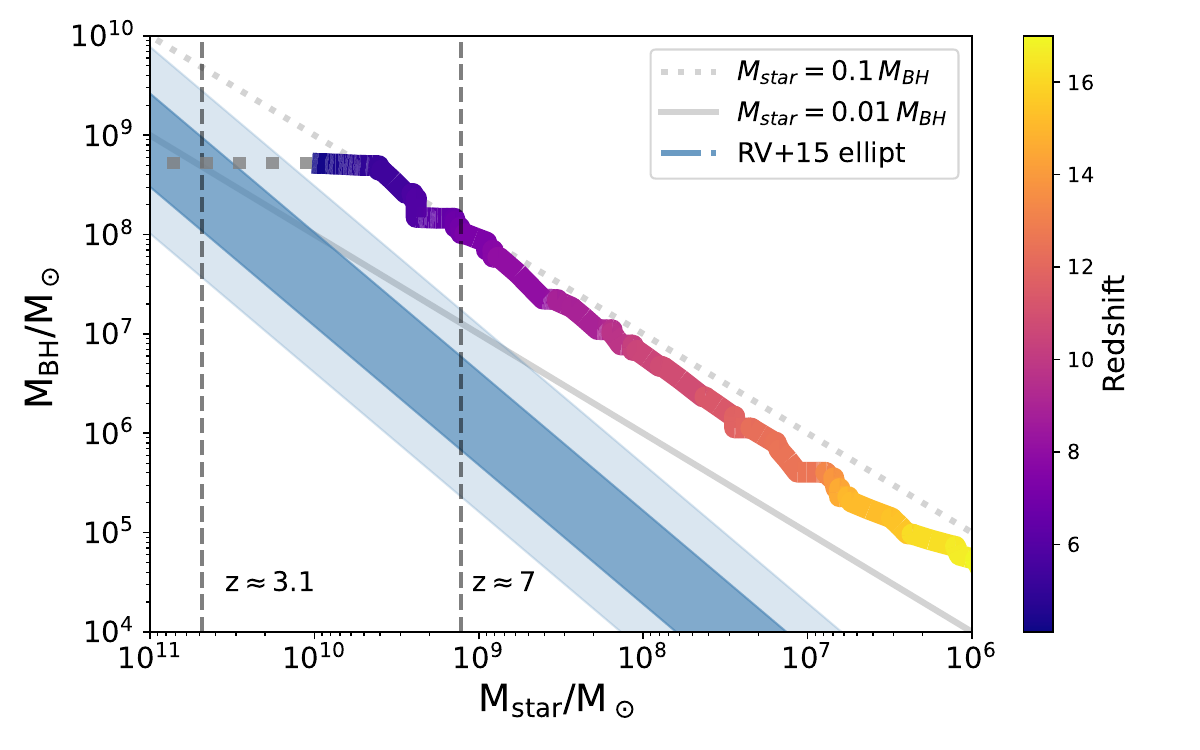}
\caption{\textit{Top panel:} Star formation history of the best synthetic counterparts of JADES GN-1001830 at $z \sim 7$ (see text). Evolutionary tracks are compared to the SFH predicted by Prospector (grey shaded area) and Bagpipes (pink shaded area) from the host galaxy SED fitting \citep{Juodzbalis2024}. For one of the candidates, we show the SFR evolution down to $z=4$. Vertical dashed lines mark the occurrence of major mergers involving the AGN host galaxy. \textit{Bottom panel}: Evolution in redshift of the $\rm M_{BH}/M_{star}$ ratio for the selected counterpart of GN-1001830 (red line in the upper panel). Dark and light blue shaded areas mark the region consistent with the local scaling relation from \citet{reines2015} within  $1 \, \sigma$ and $2 \, \sigma$, respectively. The vertical line shows the cosmic epoch when the system is expected to enter the local scaling, assuming the central MBH remains dormant.}
\label{fig:SFH}
\end{figure*}

The evolution of overmassive systems through short phases of super-Eddington accretion leads to short AGN duty cycles, of the order of few percent. This implies that a large fraction of the MBH population is expected to be “dormant”, i.e. accreting significantly below the Eddington limit, with $\lambda_{\rm Edd} = L_{\rm bol}/L_{\rm Edd} \lesssim 0.05$ \citep{schneider2023}. This scenario poses substantial challenges for the detection of these early accreting BHs, since only galaxies hosting the most massive ones would exhibit clear signatures of the nuclear source, such as a broad emission line component. For completeness, in Appendix \ref{sec:appendix} we present the distribution of Eddington ratios across the accreting massive BH population predicted by CAT in the super-Eddington model, together with its redshift evolution from $z = 10$ to $z = 5$, and discuss the comparison with recent observations.

Recently, \cite{Juodzbalis2024} reported the detection of an overmassive dormant BH hosted in the JADES GN-1001830 galaxy at $z=6.67$. The system has been covered with NIRCam and NIRSpec observations, revealing  clear broad $\rm H \alpha$ emission with a Full Width at Half Maximum (FWHM) of $\sim 5700 \, \rm km \, s^{-1}$, which leads to an estimated BH mass of $\rm log(M_{BH} / M_\odot) = 8.61 ^{+0.27} _{-0.24}$. The broad component appears to be symmetrical, and no broad emission is observed in $\rm [OIII]$, disfavouring the possibility of it being associated with an outflow. The AGN bolometric luminosity, $L_{\rm bol}\sim 3\times 10^{43} \, \rm erg s^{-1}$, estimated from the broad $\rm H \alpha$ component using the scaling relation by \cite{Stern2012}, implies an Eddington ratio $\lambda_{\rm Edd}=0.024^{+0.011}_{-0.008}$.
Finally, a stellar mass of $\rm log(M_{*}/M_\odot) = 8.92^{+0.30}_{-0.31}$ and a star formation rate of ${\rm SFR} = 1.38^{+0.92}_{-0.45} \rm \, M_\odot \, yr^{-1}$ have been inferred from the SED fitting of the 8 photometric bands using the fitting codes BAGPIPES \citep{Carnall2018} and Prospector \citep{Johnson2021}.
The dormant nature of this black hole allows for a more reliable study of its host galaxy compared to the highly accreting counterparts in the bright quasar phase, where the nuclear emission outshines the galaxy.
The MBH residing in GN-1001830 is suggested to be significantly overmassive with respect to the host galaxy, with a mass ratio $\rm M_{BH}/M_{*} \sim 0.4$, almost 3 dex above what is expected from the local scaling relation \citep{reines2015}.
As shown by \citet{Juodzbalis2024}, the properties of GN-1001830 are well explained by the \texttt{CAT} SE model, which predicts this system to be the tip of the iceberg of a vast population of dormant BHs at a similar redshift range \citep[for more details see Figure 3 of][ and the thorough discussion therein]{Juodzbalis2024}.
 
As a follow-up to that analysis, here we investigate the evolutionary tracks of simulated systems in the \texttt{CAT} SE model whose BH mass, stellar mass, and SFR are consistent, within the uncertainties, with the values inferred for GN-1001830 at $z=6.67$. Specifically, we select all systems whose three properties are simultaneously consistent with the observed values within $3\sigma$
\footnote{Note we do not impose here an explicit selection on the BH accretion rate at the observation epoch. In fact, BH accretion in \texttt{CAT} is highly episodic, making such a selection strongly dependent on the specific snapshot considered. Moreover, observed Eddington ratios are affected by detectability biases, as broad-line emission from weakly accreting systems may not be resolved (see e.g. the discussion of completeness in \citealt{Juodzbalis2024}). We note though that the selected counterparts are predominantly in a low accretion state at $z\sim7$, with $f_{\rm Edd} \lesssim  0.01$ consistent with the low Eddington ratio inferred for GN-1001830. See also Appendix \ref{sec:appendix}.}. Only 7 systems satisfy these criteria at $z \sim 7$, corresponding to a cosmological number density of $\sim 3 \times 10^{-7} \, \rm cMpc^{-3}$, which implies an expectation of finding $\lesssim 1$ such system within the cosmological volume covered by JADES. This highlights GN-1001830 as a rare object within the underlying MBH population, whose detectability is likely favoured by its large BH mass, allowing the broad emission line components to be resolved despite the low accretion rate. 
We stress that the aim of this analysis is not to characterize the average evolution of the simulated population, but rather to assess whether the model can reproduce the full set of observed properties of this extreme outlier through at least one self-consistent evolutionary history.
For the selected systems, we trace back in time the evolution of the host galaxy and its progenitors. 
In the upper panel of Figure \ref{fig:SFH} we compare the predictions of the Prospector and BAGPIPES SED fitting models for GN-1001830 with the SF histories (SFHs) predicted by \texttt{CAT} smoothed over a timescale of approximately 50 Myr to eliminate short-term variability, which is not resolved by SED fitting models. The SED fitting has been performed assuming the stellar component dominates the source emission, but excluding filter bands associated with strong AGN emission lines which might contaminate the analysis. We note the presence of a systematic offset between the SFHs inferred by the two SED fitting codes, likely arising from their different underlying assumptions regarding stellar population properties and evolutionary modeling. Nevertheless, both models consistently indicate a relatively flat SFH, with the SFR remaining in the range $\approx 0.5 - 5 \rm ~M_{\odot}/yr$ throughout most of the galaxy's evolutionary history.
In the selected \texttt{CAT} galaxies the rate of star formation slowly increases with time down to redshift $z\sim 10-12$, below which it decreases, reaching the value inferred from the observations of GN-1001830 (black square data point). Overall, the SFHs predicted by \texttt{CAT} are consistent with those inferred using BAGPIPES, within the uncertainties (pink shaded region), supporting a relatively flat SFH, with the SFR remaining almost constant, within a factor of 3, in the redshift range $z \sim 7 - 17$.
We stress that such a nearly flat SFH is not representative of the general population of overmassive BH hosts predicted by \texttt{CAT}. Indeed, the SFHs of the sample investigated in Section \ref{sec:overmassiveBHs} are broadly rising toward lower redshifts, as can already be appreciated in Figure \ref{fig:AsGrowth}, where the majority ($\sim 80\%$) of the final stellar mass of overmassive BH hosts is assembled at $z < 8$, consistent with a late and extended build-up. In the counterparts selected to match GN-1001830, the nearly constant SFR does not reflect a permanent BH-induced quenching, but rather a regulated suppression due to the repeated action of a strong episodic AGN feedback starting already at early times ($z \sim 13$), which allows the simultaneous build-up of a very massive BH and a comparatively modest stellar component. What we want to highlight here is that, while atypical for the overall population, such an evolutionary path appears to be necessary in order to match the rare and extreme properties of GN-1001830.

To explore the possible evolution of the GN-1001830 host galaxy at later cosmic epochs, in the same figure we show with a solid red line the SFH of one of the synthetic counterparts predicted by \texttt{CAT} down to $z \simeq 4$ (the final redshift of our simulation). The vertical dashed lines indicate the redshift at which this system experiences major mergers, which, in the SE model, trigger a phase of enhanced BH accretion. The star formation rate rapidly increases between major merger events, when BH accretion and feedback become less efficient. In this system, the last major merger event occurs at $z \sim 5.18$ and the SFR continuously increases thereafter becoming one order of magnitude higher by $z=4$. As a result, the stellar component grows faster than the MBH, doubling its mass and reaching a value $\rm log(M_{*}/M_\odot) \simeq 10.0$ in $\Delta t \sim \, \rm 300 ~Myr$. At $z \simeq 4$, the BH-to-stellar mass ratio is $\sim 0.05$.

The evolution in redshift of the $\rm M_{BH}/M_{star}$ ratio for the selected evolutionary track is shown in the lower panel of Figure \ref{fig:SFH}. It is clear how the system evolves along the $\rm M_{BH}/M_{star} \sim 0.1$ relation down to $z \approx 7$, when the occurrence of prolonged phases where the central BH remains 'dormant' drives the system toward lower mass ratios.
Given this result, it is interesting to provide a simple analytical estimate of the time required for the galaxy to arrive at the local $M_{\rm BH}-M_{\rm star}$ scaling relation. To this aim we assume that the galaxy does not undergo significant dynamical interactions during its subsequent evolution, so that the nuclear BH remains dormant. Considering a redshift evolution of the galaxy specific SFR as $sSFR = SFR/M_{\rm star} \propto (1+z)^{1.6}$ \citep[see][]{DiCesare2023}, we estimate that the system shown by the red line in Figure \ref{fig:SFH} would need  $\sim 0.50$ Gyr ($z\sim 3.1$) to lie within 1$\sigma$ from the local scaling relation by \citet{reines2015} for elliptical galaxies. Notably, the $\rm M_{ BH}-M_{star}$ relation for elliptical galaxies in the local Universe is steeper than what is observed for AGNs. This implies higher expected $\rm M_{ BH}/M_{star}$ ratios in massive galaxies, offering additional support for the hypothesis that these systems could align with local observations through moderate evolution of the black hole-to-stellar mass ratio.

\section{The AGN Luminosity Function at $5 < \MakeLowercase{z} < 7$} 
\label{sec:AGNlumFunc}

In \citet{trinca2022} we found that merger-driven SE accretion results in a higher number density of bright sources compared to the Eddington-limited scenario, especially in the range $L_{\rm bol} \simeq 10^{45}-10^{47} \rm \, erg/s$. It is interesting to revisit this finding in light of the large number density of AGNs identified by JWST, both as type-I/type-II AGNs \citep[see e.g.][]{harikane2023, Maiolino2024bhs, scholtz2024_AGNs} and as LRDs \citep[see e.g.][]{Matthee2024, Greene2024, kokorev2023, Akins2024}. These studies show that the number density of JWST-detected systems is between 1 to 2 dex larger than the extrapolation of the luminosity function probed by the optically and UV selected quasar population at $z \simeq 5-6$ \citep{niida2020, shen2020}, and also significantly above the estimated density from the deep X-ray observations \citep{giallongo2015, giallongo2019}.

\begin{figure}
\centering
\includegraphics[width=\hsize]{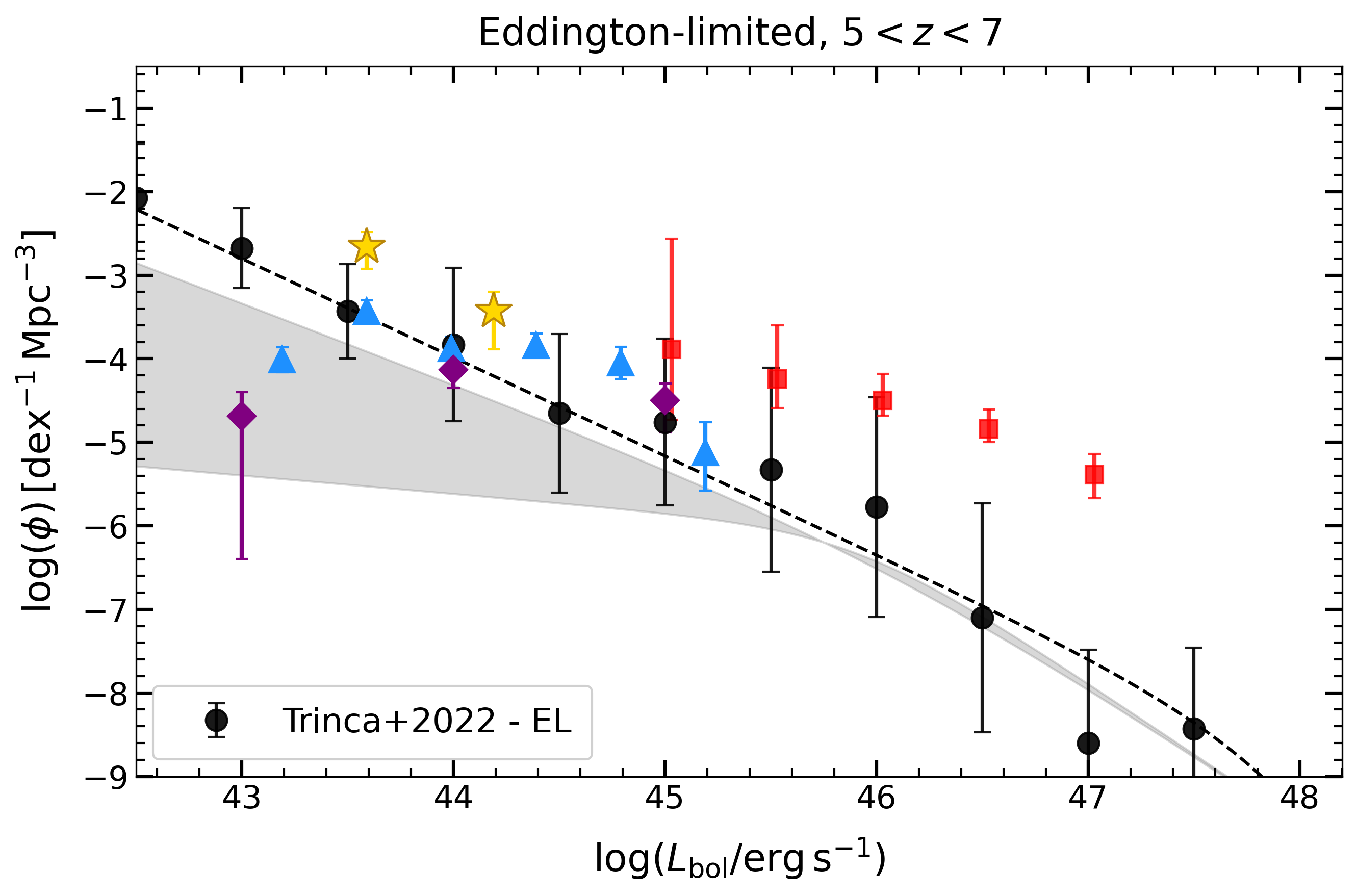}
\includegraphics[width=\hsize]{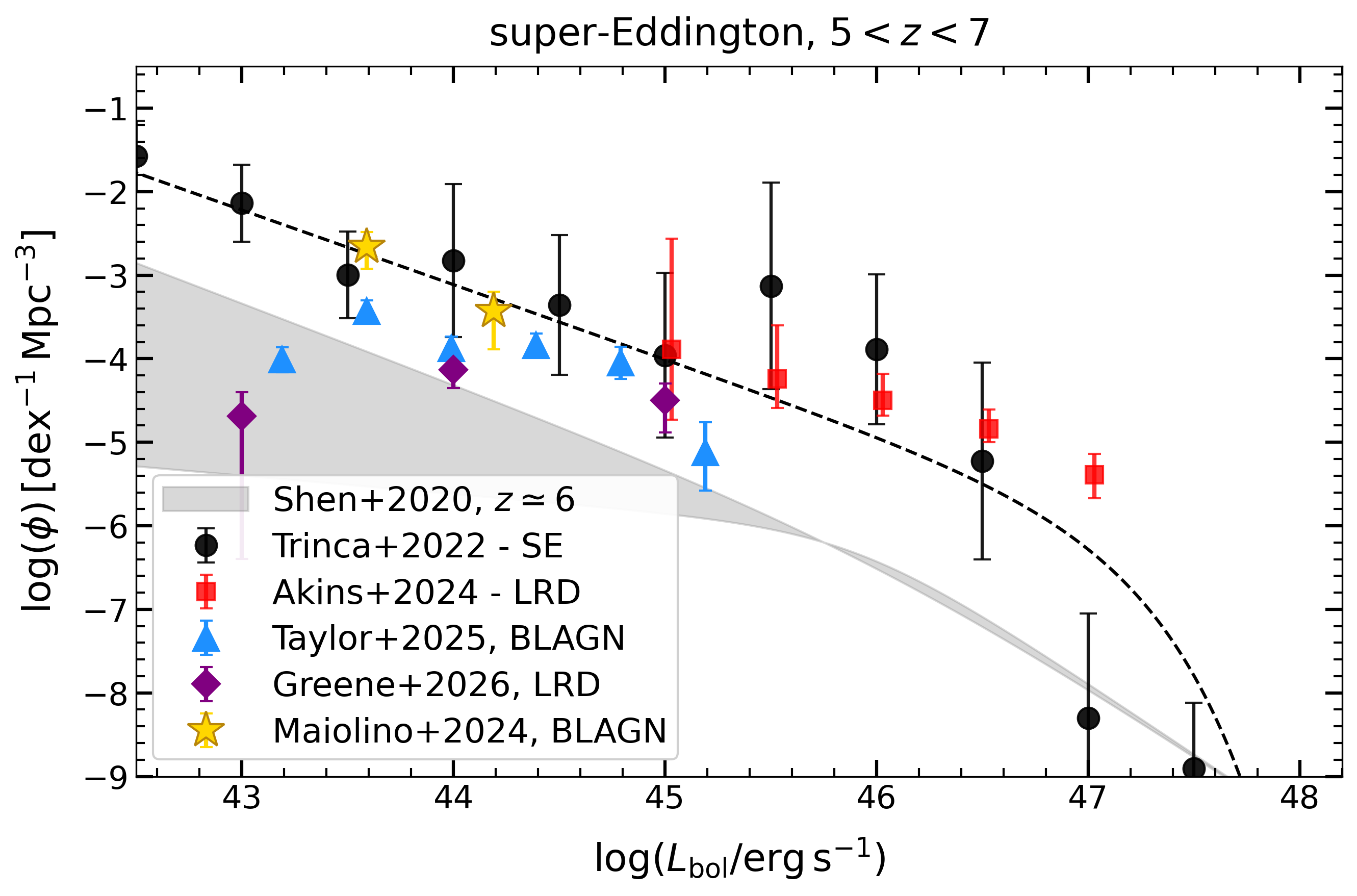}
\caption{AGN bolometric luminosity function at $5 < z < 7$. \texttt{CAT} predictions for the Eddington-limited (top panel) and super-Eddington (bottom panel) scenarios are shown as black data points, with dashed lines showing the respective best-fit distributions. These are compared with recent empirical estimates derived from JWST AGN samples. Red squares and purple diamonds show the inferred bolometric luminosity functions of 'Little Red Dots' from \citet{Akins2024} and \citet{Greene2026}, respectively. Blue triangles and yellow stars represent instead the estimates for JWST broad-line AGNs inferred by \citet{Taylor2025b} and \citet{Maiolino2024bhs}, converted from the UV luminosity function assuming standard bolometric corrections from \citet{duras2020}. The grey shaded region shows the quasar bolometric luminosity function of UV-selected quasars at $z \approx 6$ derived by \citet{shen2020}.}
\label{fig:BolLF}
\end{figure}

In Figure \ref{fig:BolLF} we show the AGN bolometric luminosity function (LF) predicted by \texttt{CAT} in the redshift range $5<z<7$ (black data points) for the Eddington-limited and super-Eddington models (top and bottom panel, respectively), together with their best-fit Schechter functions. As a comparison, we report the empirical constraints obtained for LRD in \citet{Akins2024} and \citet{Greene2026}, together with the bolometric luminosity function for the general population of JWST broad-line AGN inferred by \citet{Taylor2025b}, for $3.5 < z < 6.8$, and by \citet{Maiolino2024bhs}, for $4 < z < 6$ \footnote{The bolometric values have been inferred from the provided UV luminosity function assuming standard bolometric correction from \citet{duras2020}.}.

We can clearly see how the EL scenario predicts a bolometric distribution that lies very close to the one inferred for UV-selected quasars by \citet{shen2020} and to the related extrapolation toward the faint end. As a consequence, this model struggles to reproduce the enhanced number density of AGN inferred for the LRD population at $L_{\rm bol}=(10^{45}-10^{47})\,\rm erg\,s^{-1}$, which exceeds the model predictions by $\gtrsim 1$ dex, and, though less severely, also for the JWST broad-line AGN population, which suggests a higher normalization in the fainter regime.

Notably, the AGN bolometric LF predicted instead by the SE model in this luminosity regime is in good agreement with that derived for the LRD population by \citet[][red squares]{Akins2024}\footnote{This work analysed more than $400$ LRDs detected in the COSMOS-Web survey \citep{casey2023} at $z > 4$, and recover the bolometric luminosity from the intrinsic SED modelling, assuming that the overall emission is AGN dominated.}.
In the SE model, the predicted rise of the luminosity function above $L_{\rm bol}>10^{45} \, \rm erg/s$ results from the population of smaller-mass black holes undergoing brief episodes of SE growth, which leads to a marked boost in their luminosity. Due to the steep BH mass function, even a small fraction of active lower-mass black holes can provide a substantial contribution to the bright end of the distribution, in contrast to more massive systems undergoing a steady, sub-Eddington growth phase. This also implies that, within our modelling, a significant fraction of the systems in this luminosity range should be caught in a phase of rapid, enhanced accretion and therefore display peculiar observational features. Such a picture would be consistent with what is observed for LRDs, which show clear signatures of dense, gas-rich environments and for which super-Eddington accretion rates have been proposed \citep{Inayoshi2025, Madau2026b}.

At the same time, even in the SE scenario the predicted bolometric LF falls about one order of magnitude below the number density estimated by \citet{Akins2024} at the bright end of the distribution, for $L_{\rm bol} \gtrsim 10^{46.5}\,\rm erg/s$. It is worth noting, however, that the luminosities for this sample were estimated using standard bolometric corrections and correcting for the expected dust attenuation. Recent studies providing direct measurements of the bolometric luminosities of bright LRDs have shown that this approach may overestimate their brightness by up to an order of magnitude \citep{Greene2026}. In addition, such estimates often assume that AGN emission dominates the observed SEDs, whereas a non-negligible contribution from the host-galaxy stellar component, especially in the UV band \citep[see e.g.][]{volonteri2024, Taylor2025b}, would lower the inferred AGN luminosities, bringing them into closer agreement with those derived from mid-IR and X-ray surveys \citep{williams2023, PerezGonzalez2024, yue2024Xrays, ananna2024}. We note that, should these effects systematically lower the $L_{\rm bol}$ of the \citet{Akins2024} sample by $\sim 1$ dex, the resulting distribution would be in even better agreement with the predictions of the \texttt{CAT} SE model. We also report the more recent empirically-derived constraints on the LRD luminosity function from \citet{Greene2026}, which yield a significantly lower number density for these objects in the low-luminosity regime and are therefore less stringent in constraining the accretion regime.
Finally, tentative support for a super-Eddington scenario comes from the comparison between the \texttt{CAT} predictions and the more general bolometric luminosity function inferred by \citet{Taylor2025b} and \citet{Maiolino2024bhs} for the population of broad-line AGNs detected by JWST. The empirical data suggest a higher normalization in the fainter luminosity regimes, which lies closer to the best-fit prediction of the SE model, while exceeding the EL one by up to $\gtrsim 0.5$ dex, although both comparisons remain within the predicted model uncertainties, and are subject to uncertainties in the bolometric corrections adopted.

The observed agreement suggests that the LRD population may represent a transient evolutionary phase of early galaxies, corresponding to brief episodes of enhanced BH accretion, possibly triggered by major galaxy mergers or other significant dynamical interactions. During this phase, enhanced gas inflows onto the central BH give rise to AGN-dominated systems with the distinctive observational properties of LRDs. As suggested by \citet{Madau2026b}, a fraction of the relatively blue JWST broad-line AGNs, which share several observational features with LRDs \citep{Brazzini2026}, may likewise host super-Eddington accreting BHs, with their different spectral appearance primarily driven by viewing angle and inclination effects.

\section{Discussion}
\label{sec:discussion}

The comparison between the overmassive BH population observed at $z >4$ and the prediction of \texttt{CAT} models suggests that early BH growth could be dominated by short, repeated bursts of super-Eddington accretion. 
Interestingly, in our SE model, less than $15 \%$ of the whole population of overmassive BHs at these epochs descends from heavy seeds, thus their detection is not necessarily an indication of the existence of heavy seeds \citep[however this is not the case for earlier cosmic epochs, like $z \sim 10$, see for instance][]{natarajan2024}. In addition, we find that the typical AGN duty cycle of \texttt{CAT} SE systems is $\sim\,1-6 \%$, independent of the initial seed mass. Such short duty cycles are supported by recent clustering analysis of quasars and galaxies observed with JWST at $z\sim 6$ \citep{Pizzati2024, Eilers2024}. In our model, SE accretion is triggered during major galaxy mergers and sustained for brief periods of time (a few million years). 

The strict relation between galaxy major mergers and enhanced phases of BH accretion and star formation represents a major assumption of our modelling.
To date, observational support on whether galaxy interactions trigger enhanced phases of AGN activity is currently mixed, with studies favouring this hypothesis \citep{Goulding2018, Donley2018,Yoon2026}, while others finding no similar connection \citep{Villforth2017}. One of the main difficulties lies in the challenge of discerning a clear AGN activity during these turbulent phases, since the accreting BH might be highly gas and dust obscured even in the earliest phases of merger \citep{Kocevski2015}. However, both deep infrared observations of X-ray selected AGNs \citep{koss2018}, and radio detections of highly accreting quasars at $z \sim 5$ \citep{Trakhtenbrot2017} reported high statistical significance in  support of major mergers being important drivers for rapid black hole growth.
The recent advent of JWST also enabled us to probe in unprecedented detail the environment around bright accreting MBHs at high-redshift, resolving two massive merging systems within the host of a bright $z = 6.2$ quasar \citep{Decarli2024}.

On the numerical side, several works have investigated the role of galaxy mergers in enhancing AGN activity. Large-scale cosmological and zoom-in simulations generally find an increase in AGN activity associated with galaxy interactions, which can begin before the final coalescence and persist for extended timescales after the merger event. However, the predicted enhancement in BH accretion rates is typically moderate, with increases of up to a factor of a few \citep{Quai2022,Byrne-Mamahit2023, Byrne-Mamahit2024, Schechter2025}. Moreover, merger-related AGN enhancements have been found to be more pronounced in gas-poor systems \citep{Jhee2026}, suggesting that the impact of mergers on AGN activity may depend on the gas content and evolutionary state of the host galaxies, potentially making it less significant in gas-rich high-redshift systems. It is important to note, however, that most of these studies focus on relatively mature and massive galaxies at later cosmic epochs $0 \lesssim z \lesssim 3$, rather than on the lower stellar mass regimes and high-redshift systems probed by JWST and investigated here. Furthermore, while large-volume simulations provide statistically significant samples of merging galaxies, their spatial resolution can limit their ability to capture the small-scale gas inflows and nuclear processes that regulate the most intense phases of BH accretion.
Many hydrodynamical simulations predict that during major mergers - even when starting from disk-dominated structures - the majority of the gas is funnelled into a compact nuclear cloud on scales of a few hundred parsecs. High-resolution simulations have shown that spatial resolutions on the order of $\sim10$ pc are required to properly resolve the large gravitational torques that trigger strong nuclear inflows on smaller scales. When these are accounted for, the merger phase of galaxies often corresponds to the most sustained AGN activity \citep{capelo2015}. Recent studies also pointed out that halo collisions, especially through flybys, foster a higher-density environment that allows BH seeds to accrete at significantly higher rates than in isolated halos \citep{Prole2024}. Motivated by these evidence, we assumed that strong torques and the consequent enhancement in BH accretion and star formation occur only during galaxy major mergers.
While also secular instabilities can significantly contribute to the regulation of gas inflow toward the central black hole \citep{Garland2024}, in this study we focus on high-redshift sources at $z > 4$, where there is limited time for secular processes to develop. In these early environments, black hole growth is more likely to be dominated by strong inflows triggered by external perturbations during mergers, favoured by the compact sizes and disturbed morphology of these early galaxies.

Another caveat of our modelling is the dependence of the predicted BH accretion histories on the adopted prescription for merger-driven growth, in particular the parameters $\epsilon_{\rm BH}$ and $t_{\rm accr}$, as well as the criterion adopted to terminate the super-Eddington phase. These parameters were originally explored and calibrated in \citet{pezzulli2016} to reproduce the observed properties and evolutionary histories of luminous quasars at $z\sim6$, which constitute one of the primary observational benchmarks of our framework. As discussed in Section \ref{sec:BHgrowth}, we adopt a relatively conservative upper limit for the BH accretion timescale, while assuming the same prescription for the entire simulated galaxy population. This simplified treatment neglects possible variations of the accretion efficiency and burst duration with galaxy properties and environment. Nevertheless, our main conclusions primarily rely on the bursty nature of BH growth induced by short, merger-driven accretion episodes, while variations in the adopted parameters  will mainly affect the quantitative details of the predicted accretion histories.
Indeed, as originally shown in \citet{trinca2022}, within our modelling framework, simply increasing the BH accretion efficiency was found to be insufficient to reproduce the observed high-redshift quasar population, as well as the newly detected JWST AGNs, because the associated increase in cumulative AGN feedback at early times becomes disruptive and suppresses subsequent BH growth. In contrast, short-lived accretion bursts can enable rapid BH growth while limiting the overall impact of feedback.

Although our description of BH accretion and feedback in the super-Eddington regime is highly simplified, the predicted durations of SE accretion bursts ($\sim 1 \, \rm Myr$) are consistent with those inferred from high-resolution hydrodynamical simulations of isolated gas rich galaxies hosting BHs of $10^3$-$10^4$ M$_\odot$ \citep[e.g.][]{Lupi2016, sassano2023, massonneau2023, Gordon2024, Shi2024, Zana2026}. In these dense environments, the interaction with dense gas clumps likely enables super/hyper-Eddington accretion that is sustained for up to a few tens of million years, depending on the mechanical/kinetic feedback efficiency \citep{Lupi2024SE, Quadri2025}. Observations from JWST suggest that early galaxies are gas-rich and characterized by extremely dense and compact clumps \citep{fujimoto2025}, providing favourable conditions for super-Eddington growth \citep{Guia2024}. It has to be noted that, although major galaxy mergers have been suggested as a possible mechanism to counteract the negative effect of mechanical feedback onto BH growth and promote efficient accretion, their role in directly triggering/enhancing BH growth is still poorly explored with simulations. A shorter duration of accretion bursts onto the BH, possibly regulated by strong AGN feedback, might limit the growth of lighter BH seeds compared to our findings \citep{regan2019,Caleno2026}.
The nature of these short bursts of super-Eddington accretion, and their interplay with star formation across different galaxy and BH mass scales, will be investigated in future work using high-resolution cosmological simulations, with the aim of providing valuable insights to refine the predictions of our semi-analytical models.

An early massive BH whose growth is dominated by short, repeated phases of super-Eddington accretion has several implications for the observable features of these systems. As shown in Section~\ref{sec:AGNlumFunc}, it could help explain the large observed number density of AGNs, particularly among the LRD population, wherein it may represent a transient phase in early black hole evolution.
Interestingly, with few exceptions \citep{Kocevski2023, Bogdan2024, kovacs2024}, the AGNs detected by JWST at $z>4$ are not detected in X-rays \citep{Maiolino2024Xray, ananna2024}, and none have been found in radio observations to date \citep{Mazzolari2026}. Several explanations have been proposed, including strong AGN obscuration, differential obscuration, or intrinsically X-ray weak AGN spectra \citep{Maiolino2024Xray, ananna2024,Inayoshi2025} for these sources.
Recent theoretical and observational studies have suggested that a compelling scenario to explain many of the observed properties - at least for the LRD population - is that of AGNs embedded within extremely dense, quasi-spherical and thermalised gas envelopes, in which the central black hole may undergo a rapid, potentially super-Eddington, growth \citep[see e.g.][]{Ji2025,Inayoshi2025,deGraaff2025b}.
Regimes of super-Eddington accretion are expected to be characterized by the formation of a geometrically and optically thick accretion disk. The altered accretion geometry results in strongly collimated emission and softer X-ray spectra, characterized by a steeper photon index and a lower-energy cutoff. This effect has been proposed to arise from the overcooling of the coronal plasma embedded in a funnel-like structure \citep{Madau2024}, along with the low likelihood of observing strongly beamed X-ray emission aligned with the jet \citep{Pacucci2024grmhd,king2024}. Notably, analogous spectral features have been observed in different AGN surveys, targeting both the population of more luminous and massive quasars detected at $z \gtrsim 7$ \citep{ Zappacosta2020, Zappacosta2023} and samples of more local supermassive BHs accreting at super-Eddington rates \citep{Tortosa2024}. However, the full multi-wavelength signatures of SE accretion remain currently unsettled. Although a handful of sources with implicated SE accretion have been identified in the redshift range explored in this work, there is considerable variation in their detected rest-frame UV-optical and X-ray fluxes. For instance, the source LID-568 \citep[z = 4.4, ][]{suh2024} appears X-ray bright, while other systems, like PSO J006+39 \citep[z = 6.6, ][]{Tang2019}, exhibit signatures of highly obscured, geometrically thick accretion disks.

Meanwhile, high Eddington ratios could also mitigate the tension between the observed overmassive AGNs detected by JWST and the local relation. Indeed, \citet{Lupi2024} showed that, as a consequence of the modified disk geometry, for high Eddington ratios the self-shadowing effect of the accretion disk may lead to overestimating the BH masses by up to an order of magnitude when relying on the standard, single-epoch virial methods, based on local reverberation mapping calibrations. Recent direct, interferometric measurement of the BLR in a distant quasar accreting highly super-Eddington revealed an offset ranging between $0.43-0.73$ dex, depending on the chosen calibrator \citep{Abuter2024}.

An additional concern has been raised by \citet{Inayoshi2024b}. They show that, under the assumption of an AGN-dominated emission and a population spanning the wide redshift range where LRDs are detected ($4 \lesssim z \lesssim 8$, \citealt{Kocevski2024}), the BH mass density accreted by LRDs at $z > 4$ would already exceed the local BH mass density. While they invoke the need for high radiative efficiencies during early BH accretion, suggesting higher BH spin, this tension might be similarly alleviated assuming a growth history marked by very short-lived phases of efficient BH accretion - significantly shorter than the observational time window covered by the LRD sample, of the order of $\sim 1$~Myr as we report here. 

Interestingly, \citet{Mezcua2024} recently reported the detection of 12 overmassive BHs in low-mass galaxies observed at cosmic noon ($z \sim 1-3$) in the VIMOS Public Extragalactic Redshift Survey (VIPERS). These galaxies present physical properties - such as BH mass, stellar mass, bolometric luminosity, Eddington ratio - that are remarkably similar to the ones inferred for the high-$z$ AGNs observed by JWST, suggesting that we might be observing systems in an analogous phase of their evolution, but during different cosmic epochs. If we assume that galaxies hosting overmassive BHs transition to the local scaling relations on timescales comparable to those estimated by \texttt{CAT} ($\sim 0.5-1 ~\rm Gyrs$), this could explain why systems with such extreme $\rm M_{\rm BH}/M_{\rm star}$ ratios may not have been observed at $z \simeq 0$.

Connecting the early overmassive BH population discussed in this work with the present-day galaxy population will require a future extension of the CAT model towards lower redshifts, accounting for the increasing importance of processes such as secular gas inflows and morphological transformations, while simultaneously improving the self-consistent modelling of galaxy and BH evolution at the earliest cosmic epochs, where JWST has revealed significant departures from pre-existing theoretical predictions. In particular, this will require modelling the evolution of galaxy properties under the rapidly changing physical conditions of the high-redshift Universe \citep[see, e.g.,][]{somerville2026}, as well as refining the treatment of super-Eddington accretion episodes through improved prescriptions calibrated against dedicated high-resolution simulations \citep[e.g.][]{Zana2026,Caleno2026}. Such developments will enable us to investigate the coupled evolution of galaxies and massive BHs from the first billion years of cosmic history to the local Universe, and to provide a more complete interpretation of the emerging observational constraints.

\section{Conclusions} 
\label{sec:conclusions}
In this study, we employed the Cosmic Archaeology Tool semi-analytical model, originally presented in \citet{trinca2022}, to interpret the growing JWST sample of high-redshift AGNs, which now extends toward lower luminosities, allowing us to probe the faint end of the population. This enables us to assess whether their properties, and in particular their large BH-to-stellar mass ratios inferred, can be used to constrain the dominant underlying mechanism of black hole growth.
We show that the AGNs observed by JWST at $4<z<7$ favour a scenario where light and heavy BH seeds grow primarily through short phases of super-Eddington accretion. In the CAT model, these phases are triggered by galaxy major mergers. We reconstruct the evolutionary pathways of overmassive BHs at $z \gtrsim 5$, investigating how they accumulate mass over cosmic time and interact with their host galaxies.

We find that:
\begin{itemize}
\item Overmassive BHs with properties consistent with JWST observations can originate from both light and heavy seeds. In particular, the vast majority ($> 85 \%$) of systems predicted by \texttt{CAT} showing analogous properties descend from light-PopIII remnant seeds formed at $z>20$, while only a minority originates from direct collapse BHs formed between $z \sim 13-15$. The highly efficient episodes of accretion rapidly erase the information on the nature of the BH seed, making it difficult to differentiate between these formation pathways for individual sources by $z \sim 8$.\\

\item The selected galaxy population shows a common characteristic evolutionary history, independent of the BH seeding channel. At $z>8$, the stellar component and the main MBH progenitor face a decoupled evolution, growing in different halos as a consequence of the efficient feedback exerted by the rapidly growing BH on its host galaxy. The MBH-galaxy co-evolution starts only around $z \sim 8$, when the galaxy has assembled, on average, $30 \%$ of its final stellar mass. In the subsequent evolutionary phase, the impact of MBH feedback on the galaxy decreases in importance as a result of the decreasing number of super-Eddington events and the growth of the halo's gravitational potential.\\

\item Our model predicts relatively short evolutionary timescales and duty cycles associated with the formation of overmassive systems, even when starting from smaller BH seeds. We find that single episodes of super-Eddington accretion typically last between $0.5-3$ Myr, with a median value of $\sim 1$ Myr, resulting in a total AGN duty cycle of $\sim 1-4 \%$. No clear correlation is observed between the growth timescale and the nature of the initial seed, confirming that BH growth is primarily governed by their evolutionary environment. These short duty cycles suggest that a large fraction of the MBH population at high redshift is in an inactive phase, as supported by recent observations, including the dormant overmassive BH detected in the JADES GN-1001830 galaxy at $z \sim 6.67$.\\

\item We investigated whether the \texttt{CAT} SE model can reproduce the full set of properties inferred for the overmassive BH in JADES GN-1001830, representing an extreme outlier even within the observed overmassive population. We find a small number of synthetic counterparts simultaneously matching its BH mass, stellar mass and SFR, within current uncertainties. These systems share a nearly flat SFH, consistent with what is inferred from SED fitting, likely resulting from strong, episodic AGN feedback regulating star formation from early times, which is required to build up a very massive BH alongside a comparatively modest stellar component. Tracking the subsequent evolution down to $z=4$, we find that, as the frequency of major mergers decreases, BH growth slows down and star formation rapidly rises, increasing the galaxy stellar mass by $\sim 1$ dex. This transition drives these systems toward lower BH-to-stellar mass ratios, possibly allowing them to reach the local scaling relations in $\sim 0.5$ Gyr, assuming the MBH remains dormant during this time.\\

\item The enhancement in BH luminosity associated with short periods of super-Eddington accretion results in a bolometric AGN luminosity function at $5 < z < 7$ well aligned with recent JWST observations. Specifically, the \texttt{CAT} super-Eddington model closely reproduces the high number density of accreting BHs with $ \rm L_{bol}= 10^{45} - 10^{47} \, erg/s$, as inferred from the population of “Little Red Dots”, under the assumption that their emission is primarily AGN-dominated, and is also supported, though more tentatively, by the large number densities inferred from broad-line AGN samples. This supports the idea that the LRD population might represent a transient phase in early galaxy evolution.\\
\end{itemize}

\section*{Acknowledgments}
 AT sincerely thanks the Kavli Institute for Cosmology, Cambridge (KICC) for their extended hospitality that greatly influenced and supported the development of this work. This research was supported in part by grant NSF PHY-2309135 to the Kavli Institute for Theoretical Physics (KITP). AT, RS and RV thank the KITP and the organisers of the “Cosmic Origins: The First Billion Years” program for useful discussions and their hospitality during the program, which significantly contributed to the development of this work. AT, RV, and AL acknowledge support from PRIN MUR “2022935STW” funded by European Union-Next Generation EU, Missione 4 Componente 2. AT, RS and RV acknowledge financial support from the Bando Ricerca Fondamentale INAF 2023, Theory Grant “Theoretical models for Black Holes Archaeology” and Mini-grant “Cosmic Archaeology with the first black hole seeds”. RV acknowledges financial support from the Bando Ricerca Fondamentale INAF 2022 Large Grant “Toward an holistic view of the Titans: multi-band observations of $z > 6$ QSOs powered by greedy supermassive black-holes”. RS, LG, and TZ acknowledge support from the PRIN 2022 MUR project 2022CB3PJ3—First Light And Galaxy aSsembly (FLAGS) funded by the European Union—Next Generation EU, and from EU-Recovery Fund PNRR - National Centre for HPC, Big Data and Quantum Computing. LG acknowledges support from the Amaldi Research Center funded by the MIUR program “Dipartimento di Eccellenza” (CUP:B81I18001170001). PN acknowledges support from the Gordon and Betty Moore Foundation and the John Templeton Foundation that fund the Black Hole Initiative (BHI) at Harvard University where she serves as an external PI. RM acknowledges support by the Science and Technology Facilities Council (STFC), by the ERC through Advanced Grant 695671 ”QUENCH”, and by the UKRI Frontier Research grant RISEandFALL. RM also acknowledges funding from a research professorship from the Royal Society.

\section*{Data Availability}
The simulated data underlying this article will be shared on reasonable request to the corresponding author.


\bibliographystyle{mnras}
\bibliography{overmassiveBHs} 

\appendix
\section{Black hole demographics}
\label{sec:appendix}
\begin{figure}
\centering
\includegraphics[width=\hsize]{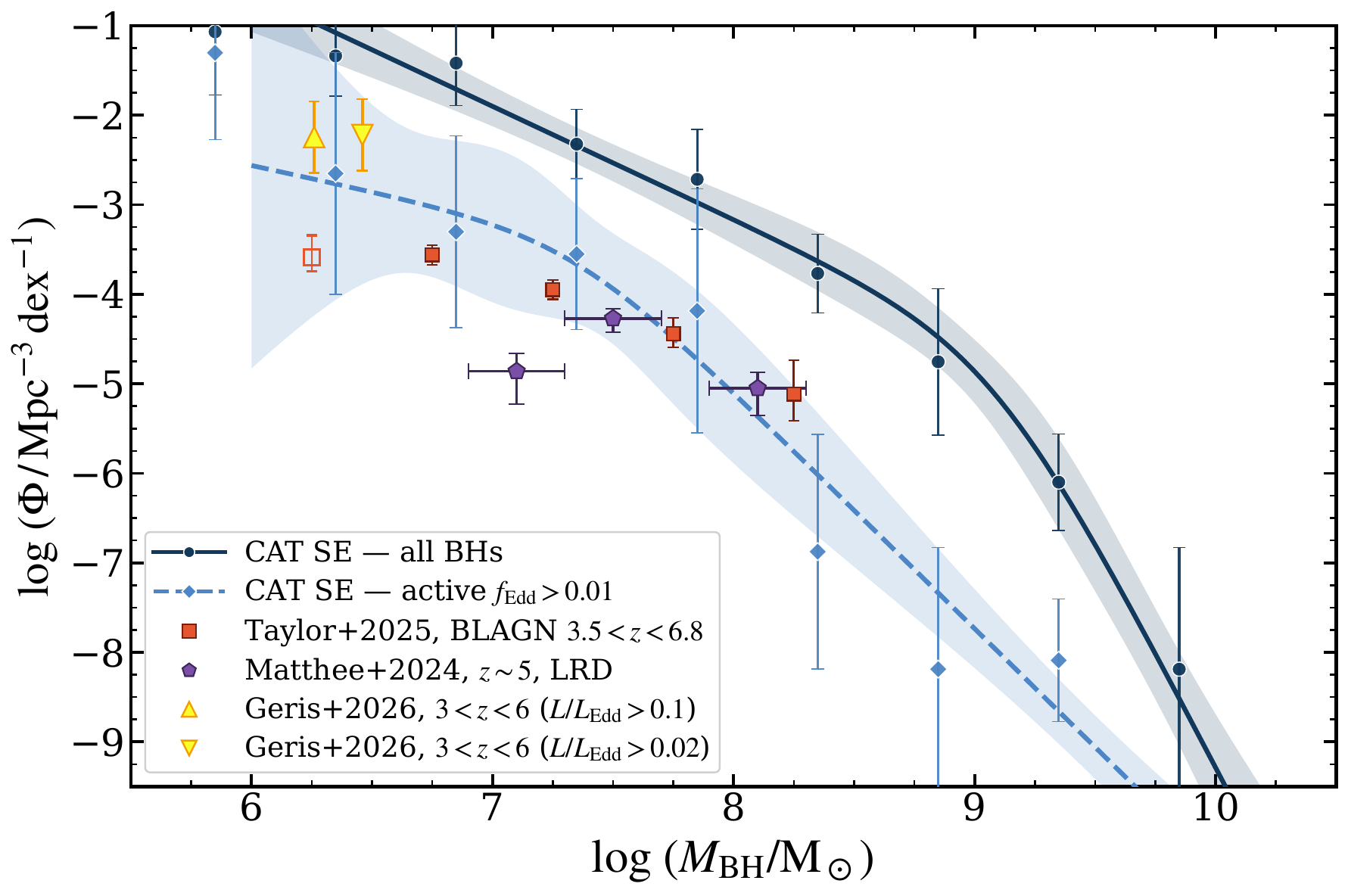}
\caption{Comparison between the predicted black hole mass function and recent observational constraints from JWST at the low-mass end. Black data points represent the predicted BH mass function at $z \sim 5$ for the CAT super-Eddington model, while blue data points represent the same distribution, but accounting only for active systems with $f_{\rm Edd}>0.01$. For both distributions, best-fit Schechter functions with related uncertainty are shown with the same colours. Observational constraints from \citet{Matthee2024,Taylor2025b, Geris2026} are reported, covering the mass range $10^6 \lesssim \rm{M_{\rm BH}/M_\odot} \lesssim 10^8$. These are all obtained from the detection of broad emission lines, tracing efficiently accreting systems.}
\label{fig:BHMF}
\end{figure}

Here we provide additional diagnostics for the CAT super-Eddington scenario, comparing the predicted black hole mass function (BHMF) and Eddington ratio distribution with recent observational constraints obtained from independent JWST surveys and samples. In Figure \ref{fig:BHMF}, we show the predicted BHMF at $z \sim 5$. CAT predictions for the total BH population and the active AGN population (defined as objects with $\lambda_{\rm Edd}=L/L_{\rm Edd}>0.01$) are shown as black and blue datapoints, respectively. The corresponding best-fitting Schechter functions are shown with solid and dashed lines, together with their uncertainties (shaded regions).

These theoretical predictions are compared with empirical constraints derived from JWST observations. In particular, we show the BHMF inferred by \citet{Taylor2025b} from the analysis of 62 broad-line AGNs at $3.5 < z < 6.8$ identified in the CEERS and RUBIES surveys through their broad $H\alpha$ emission. We find very good agreement between the CAT predictions and these observational estimates over the range $M_{\rm BH}\sim10^{6.5}-10^8 \rm ~M_\odot$. At lower BH masses, the model predicts a larger number density of objects; however, this regime is also where the authors report decreasing detection completeness, making a direct comparison more uncertain.

We further compare our results with the deeper constraints proposed by \citet{Geris2026}, based on the detection of broad $H\alpha$ components through stacking of 600 NIRSpec spectra from the JADES survey. This analysis extends the observational constraints towards lower masses, probing the range $6.0 < \log(M_{\rm BH}/M_\odot) < 6.5$, and is expected to be sensitive to BHs accreting at $\lambda_{\rm Edd}\gtrsim0.02$. The inferred number densities, reaching values as high as $\sim6 \times10^{-3} ~{\rm Mpc}^{-3}$ for low-mass MBHs, are consistent with the predictions and uncertainties of the CAT super-Eddington model.

For completeness, we also report the BHMF originally derived for LRDs by \citet{Matthee2024}, which suggests number densities comparable to those of broad-line AGNs at $7.5 < \log(M_{\rm BH}/M_\odot) < 8.5$. However, these estimates should be taken with caution, as recent studies suggest that the masses of LRDs may be systematically overestimated by up to, or even more than, one order of magnitude. Possible contributions include the application of standard bolometric corrections to their UV emission, which may significantly overestimate their intrinsic luminosities \citep{Greene2026}, the contribution of electron scattering to the broadening of Balmer emission lines \citep{Rusakov2026}, and systematic biases in single-epoch virial mass calibrations when applied to sources accreting close to or above the Eddington limit \citep{Lupi2024, Trinca2026}.

\begin{figure}
\centering
\includegraphics[width=\hsize]{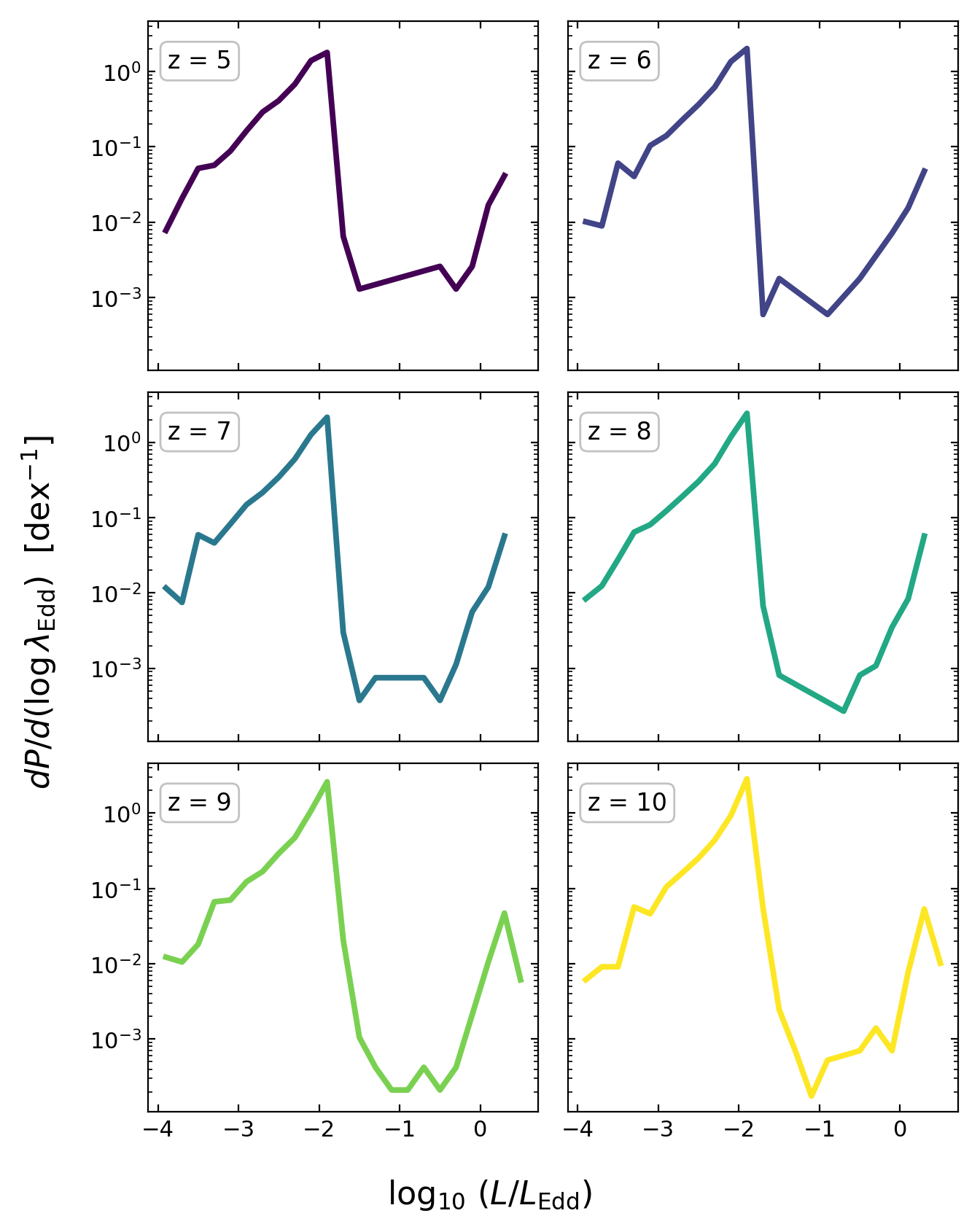}
\caption{Probability density distribution of the bolometric Eddington luminosity ratio, $\log_{10}(L/L_{\rm Edd})$, for accreting BHs in the CAT super-Eddington model at redshifts $z=5-10$. Each panel shows the normalised probability density per dex at the corresponding redshift. The distribution exhibits a bimodal shape, reflecting the different accretion regimes predicted by the model, where a high-$\lambda_{\rm Edd}$ component arises from episodes of enhanced accretion following galaxy major mergers. This component shows a mild evolution towards lower redshifts, with a gradual reduction in the maximum Eddington ratios reached by the population.}
\label{fig:EddRatios}
\end{figure}

In Figure \ref{fig:EddRatios}, we show instead the probability density distribution of Eddington luminosity ratios predicted by the CAT super-Eddington model at different redshifts, from $z=10$ to $z=5$. The distribution exhibits a clear bimodal shape, reflecting the different accretion regimes experienced by the BH population. In particular, efficient growth occurs preferentially either in more massive, gas-rich systems, or during short phases of enhanced accretion triggered by major merger events, which can drive BHs into the super-Eddington regime.

We observe a mild evolution of the Eddington ratio distribution with redshift. At earlier epochs ($z\sim9-10$), BHs are able to reach higher values of $\lambda_{\rm Edd}$, as a consequence of their lower typical masses and the availability of abundant gas reservoirs, which allow strong inflows to translate into higher accretion rates. Towards lower redshifts, the distribution progressively shifts towards lower Eddington ratios, with an increasing fraction of systems experiencing sub-Eddington accretion in the range $-1.5 < \log(\lambda_{\rm Edd}) < 0$. This evolution is naturally expected as BHs grow in mass, while the average gas supply available within their host galaxies gradually decreases over cosmic time.

Interestingly, this bimodality is qualitatively consistent with the observational picture emerging from JWST, dominated by bright, actively accreting BHs of relatively lower mass, but now including a few more massive systems caught in a weakly accreting, 'dormant' phase, such as JADES GN-1001830 \citep{Juodzbalis2024} and, more recently, Abell 2744-QSO1, whose BH mass has been inferred from direct dynamical measurements \citep{Juodzbalis2026}. While the observed population is clearly shaped by selection effects, as weakly accreting BHs are detectable only when their large masses allow the broad-line emission to be resolved despite the low Eddington ratios \citep[see e.g. the discussion in][]{Juodzbalis2024}, these few candidates may represent the tip of the iceberg of a much larger underlying population of inactive massive BHs, as predicted by our model.

\bsp	
\label{lastpage}
\end{document}